\documentclass[a4paper, amsfonts, amssymb, amsmath, prd, showkeys, nofootinbib, twocolumn, superscriptaddress, colorlinks, linkcolor=refcol, citecolor=refcol, urlcolor=refcol]{revtex4-1}
\usepackage[english]{babel}
\usepackage[utf8]{inputenc}
\usepackage{orcidlink}    
\usepackage[colorinlistoftodos, color=green!40, prependcaption]{todonotes}
\usepackage{bbold}
\usepackage{physics}
\usepackage{makecell}
\definecolor{refcol}{RGB}{178,34,34}
\usepackage{microtype}
\usepackage{tabularx}
\graphicspath{{./figures/}}
\newcommand{\be}{\begin{equation}}
\newcommand{\ee}{\end{equation}}
\newcommand{\bea}{\begin{align}}
\newcommand{\eea}{\end{align}}
\newcommand{\eff}{\mathrm{eff}}
\newcommand{\fo}{\mathrm{fo}}
\newcommand{\pc}{\mathrm{pc}}
\newcommand{\cl}{\mathrm{cl}}

\newcommand{\Tau}{\tilde{{T}}}
\newcommand{\CEP}{\mathrm{CEP}}
\newcommand{\MF}{\mathrm{MF}}
\definecolor{red}{rgb}{1,0,0}
\newcommand{\red}[1]{{\color{red} #1}} 
\definecolor{internationalkleinblue}{rgb}{0.0, 0.18, 0.65}

\def\Eq#1{Eq.~(\ref{#1})}

\def\Fig#1{Fig.~\ref{#1}}
\def\Tab#1{Tab.~\ref{#1}}

\def\Sec#1{Sec.~\ref{#1}}

\allowdisplaybreaks

\begin{document}
\title{Finite-size effects and scaling properties of chiral and baryon-number fluctuations}

\author{Győző Kovács \orcidlink{0000-0003-0570-3621}}
\email{gyozo.kovacs@uwr.edu.pl}
\thanks{corresponding author}
    \affiliation{Institute of Theoretical Physics, University of Wroc\l{}aw, plac Maksa Borna 9, PL-50204 Wroc\l{}aw, Poland}
    \affiliation{Institute for Particle and Nuclear Physics,
    HUN-REN Wigner Research Centre for Physics, 1121 Budapest, Konkoly–Thege Miklós út 29-33, Hungary}

\author{Pok Man Lo \orcidlink{0000-0002-2866-7996}}
\email{pokman.lo@uwr.edu.pl}
    \affiliation{Institute of Theoretical Physics, University of Wroc\l{}aw, plac Maksa Borna 9, PL-50204 Wroc\l{}aw, Poland}

\author{Krzysztof Redlich \orcidlink{0000-0002-2629-1710}}
\email{krzysztof.redlich@uwr.edu.pl}
    \affiliation{Institute of Theoretical Physics, University of Wroc\l{}aw, plac Maksa Borna 9, PL-50204 Wroc\l{}aw, Poland}
    \affiliation{Polish Academy of Sciences PAN, Podwale 75, 
    PL-50449 Wroc\l{}aw, Poland}
    
\author{Chihiro Sasaki \orcidlink{0000-0003-4612-3375}}
\email{chihiro.sasaki@uwr.edu.pl}
    \affiliation{Institute of Theoretical Physics, University of Wroc\l{}aw, plac Maksa Borna 9, PL-50204 Wroc\l{}aw, Poland}
    \affiliation{International Institute for Sustainability with Knotted Chiral Meta Matter (WPI-SKCM$^2$), Hiroshima University, 1-3-1 Kagamiyama, 739-8531, Higashi-Hiroshima, Hiroshima, Japan}

\begin{abstract}
    An effective chiral model is introduced to illustrate finite-size effects, incorporating the standard zero-mode treatment, momentum-space discretization, and gradient effects modeled via a prescribed finite-volume profile.
    The fluctuations of the chiral order parameter and the net-baryon number, as well as their scaling properties, are investigated near a critical point and a first-order transition in finite-size systems. 
    The finite-volume effects on the Binder cumulant and the kurtosis are shown along the phase boundary and the approximate freeze-out line, respectively. 
    Phenomenological implications on fluctuation observables are pointed out and explained.
\end{abstract}

\date{April 2026}

\hypersetup{
pdftitle={Finite-size effects and scaling properties of charge fluctuations},
pdfauthor={Gyozo Kovacs}
}

\maketitle

\section{Introduction} \label{sec:intro}

Locating the critical endpoint (CEP) in the phase diagram of strongly interacting matter is one of the main goals of recent explorations in heavy-ion physics \cite{Stephanov:2004wx, Fukushima:2013rx, Luo:2017faz, Bzdak:2019pkr}. From the theoretical side, the critical behaviors at high chemical potential or baryon density can be explored using effective models formulated in the thermodynamic limit. Experimentally, the high-density region is probed with low-energy heavy-ion collisions, where the system reaches a few or a few tens of fermi linear extent in the fireball. Due to the finite system size -- together with other effects, such as critical slowing down or finite time -- no true criticality can appear in experiments \cite{Binder:1981sa, Binder:1984llk, Binder:1985xkp, Binder:1992pz, Privman:1983, Zinn-Justin:1989rgp}. 
To observe finite-size rounding in theoretical calculations, one must either incorporate the finite-size effects within the models originally defined in the thermodynamic limit or employ frameworks that are inherently formulated in finite volumes. 

However, the theoretical approaches that are naturally suited for studying the finite-size effects are typically not directly applicable at large chemical potentials, where a critical endpoint might be found. The most prominent example is lattice QCD. Although lattice calculations provide information from the first principles on the phase transition at finite temperature ($T$) and vanishing baryon chemical potential ($\mu_B$), their extension to finite $\mu_B$ is limited due to the infamous sign problem \cite{Langelage:2009jb, Vovchenko:2018zgt}.
Various techniques have been invented to overcome this issue \cite{Borsanyi:2012ve, Karsch:2003jg, Bazavov:2019www, deForcrand:2002pa, Parisi:1983mgm, Klauder:1983zm},
and the system size dependence of the phase boundary at $\mu_B\neq0$ has also been studied on the lattice through extrapolation from imaginary chemical potentials \cite{RubenKara:2024krv, Borsanyi:2025lim}. 
Finite-size effects can also be implemented systematically in chiral perturbation theory (ChPT). This provides information on the size dependence of physical quantities, e.g., meson masses, decay widths \cite{Colangelo:2002hy, Colangelo:2003hf, Colangelo:2005gd}, and the quark condensate \cite{Gasser:1986vb, Gasser:1987ah, Bijnens:2006ve, Damgaard:2008zs, Adhikari:2023fdl}, although the results are valid only at low temperatures and vanishing chemical potentials. 
Reliable modeling of finite-volume systems valid at high density is mandatory to capture any signatures of a potential critical point expected in this domain.

In this work, we focus on the finite-size effects on the phase structure and charge fluctuations in an effective model that exhibits a critical endpoint in the thermodynamic limit. The volume dependence of the QCD phase diagram and the hypothetical critical endpoint has already been widely studied using such approaches via momentum-space constraints \cite{Palhares:2009tf, Abreu:2015jya, Abreu:2017lxf, Luecker:2009bs, Almasi:2016zqf, Tripolt:2013zfa, Bernhardt:2021iql, Magdy:2015eda, Kovacs:2023kcn, Kovacs:2023kbv}. However, when these constraints are applied to a mean-field model, the divergences at the critical point cannot be removed, and the finite-size scaling is trivially absent. 

Here, we consider a formulation with a volume-dependent partition function including zero-mode integration and derive observables in close analogy to the grand-canonical ensemble. We employ a model that exhibits a critical divergence in the infinite-volume limit, which is then regulated at finite volume, giving rise to finite-size scaling.

At the most basic level, we isolate the volume effect from the zero-mode treatment by retaining only the statistical weighting of the spatially constant (zero) mode, while using a size-independent mean-field effective potential. This corresponds to neglecting momentum-space constraints, vacuum fluctuations, and the gradient terms.

We then extend this setup by systematically incorporating additional sources of finite-size dependence. First, momentum discretization introduces an intrinsic volume dependence in the effective potential, whose effects are previously explored~\cite{Kovacs:2023kbv}. Second, gradient terms generate spatial correlations beyond the zero-mode approximation. A complete treatment of these contributions requires solving the equation of motion for the classical field and determining its spatial profile, as explored in recent works~\cite{Li:2021qcb, Lo:2026xuc}. 
Here, we instead incorporate gradient effects through a parametrized spatial profile for the mean field. This allows us to assess the impact of kinetic terms without solving the full equation of motion for spatially resolved field configurations.

We study specifically the scaling behavior and finite-size effects on various charge susceptibilities and their ratios. 
The importance of these quantities is given by their accessibility in both theory and experiments, and their expected size (in)dependence, which make them one of the main tools in the search for the CEP \cite{PhysRevLett.126.092301, STAR:2022etb, HADES:2020wpc, Kitazawa:2012at, Sorensen:2024mry}. 

Our aim is solely to provide a qualitative picture of how the transition is affected when the system has a finite spatial extent from a theoretical point of view. For the emerging phenomena, we distinguish two regimes at finite sizes: First, for a finite but large volume, the non-critical quantities are almost unaffected (since dominated by the regular part), especially at low chemical potential, while the fluctuations in the vicinity of the critical point follow the expected finite-size scaling.
Second, when the system size is sufficiently small (in the present work, at linear sizes $L\lesssim10-20$ fm, for higher-order fluctuations on the freeze-out line $L\lesssim 40$ fm), a correction to the scaling behavior emerges, while the thermodynamics and the phase diagram change significantly. These are the genuine finite-size effects. The momentum space discretization and the gradient term, becoming relevant below $L\approx10$ fm, also contribute to these effects. While there are further limitations---such as canonical effects, critical slowing down, finite time, and also the fluctuation of the finite size---in a realistic physical system, these are beyond the scope of the present work.

The paper is organized as follows. In Sec.~\ref{sec:method}, we introduce our framework to study finite-size effects in an effective model, based on the integration over a constant field mode.
We also present our model setup, as well as the technical details of parametrization and scaling. Our results for the size dependence of the different physical quantities, their scaling behavior, and the phase diagram are given in Sec.~\ref{sec:results}, where we also discuss the effect of the phase coexistence. Sec.~\ref{sec:Fluct} is devoted to the finite-size effects on the fluctuations of the net baryon number along the phase boundary and the freeze-out line. Our conclusions are given in Sec.~\ref{sec:conclusion}. An appendix is included to provide the definitions used in our framework.

\section{Effective model at finite size} \label{sec:method}

\subsection{Framework and Approximations}

We propose an approach to implement finite-size effects, including finite-size rounding and scaling.
We work with a cubic volume $V=L^d$ with $d=3$ spatial dimensions, and $L$ is the linear extent of the system.
For simplicity, we consider the case with a single scalar field $\phi(x)$. Since we aim to discuss finite temperature behavior, we start with the partition function, 
\be \label{Eq:Z_euc_def}
\mathcal{Z}=\int \mathcal{D} \phi ~ e^{-S[\phi]}\,,
\ee 
using the Euclidean action, $S[\phi]=\int_x \mathcal{L} (\phi)$ with $\int_x=\int d\tau \int d^dx$, and define the free energy, as
\be \label{Eq:freeenergy_def}
\Phi = -T \ln \mathcal{Z} = -T \ln \int \mathcal{D}\phi ~e^{-S[\phi]} \,.
\ee 
The physical quantities can be derived analogously to thermodynamics in the grand canonical ensemble, as discussed in detail in Appendix~\ref{sec:defs}. Generally, the expectation value of a physical quantity $\mathcal{A}$ is given by
\be \label{Eq:phi_def}
\langle \mathcal{A} \rangle = \int \mathcal{D} \phi ~ \mathcal{A} \, P[\phi],\qquad P[\phi]=\frac{1}{\mathcal{Z}}e^{-S[\phi]} \, .
\ee 
While carrying out the full functional integral is not possible for a general system, there are several approximations to obtain an analytically and numerically controllable theory. 

\begin{figure}
    \centering
    \includegraphics[width=0.7\linewidth]{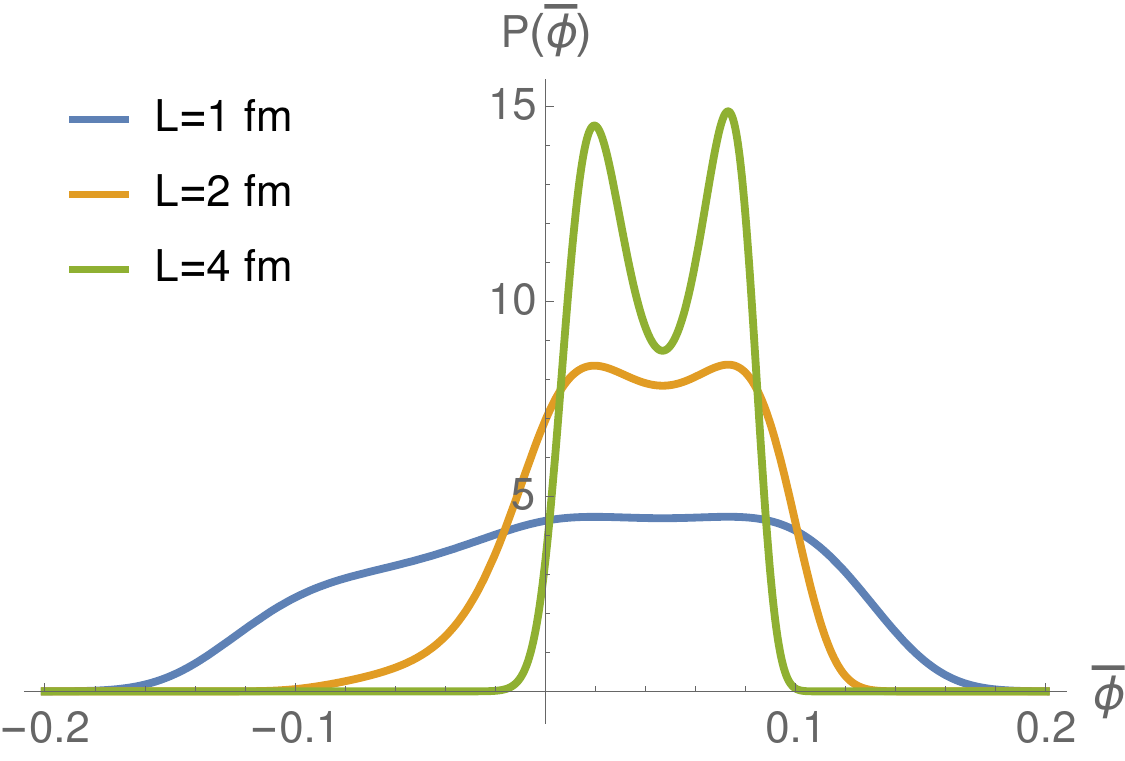}
    \caption{The $P(\phi)=e^{-\beta V\mathcal{U}_\eff(\phi)}/\mathcal{Z}$ weights as a function of the field $\phi$ for different (rather small) system sizes close below the first-order phase boundary.} 
    \label{Fig:weights}
\end{figure}

Here, we aim to utilize a framework where the integration over the infrared sector is performed nonperturbatively, which provides a natural basis to study the finite-size effects and the restoration of symmetries in finite volumes \cite{Gasser:1987ah, Gockeler:1990zn}.
To this end, we separate the constant zero mode of the field $\phi(x)=\phi_0+\delta\phi(x)$, where $\phi_0=\int d^dx \phi(x)/V$, which can be done formally without discarding degrees of freedom.
However, since the action generally couples the zero and nonzero modes, evaluating the corresponding integrals independently requires additional approximations.
In the simplest realization of such an approach, the fluctuating part $\delta\phi(x)$ is completely omitted. Therefore, only the integration over the spacetime-independent $\bar\phi\equiv\phi_0$ mode remains, and hence the partition function reads
\be  \label{Eq:Ueff_MF}
\mathcal{Z} = \int_{-\infty}^{\infty} d \bar\phi~ e^{- \beta V \mathcal{U}_\eff^{(0)} (\bar\phi)}\,.
\ee 
Here we defined the mean-field effective potential $\mathcal{U}_\eff^{}(\bar\phi)=S_E(\phi=\bar\phi)/\beta V$, where $\beta=1/T$ is the inverse temperature. 
In this approximation, any contribution from the kinetic term for $\phi$ is neglected, while assuming that the integration is still dominated by the minimum configuration. The potential may still have nontrivial $L$-dependence (e.g., from other, integrated out degrees of freedom). The form in \Eq{Eq:Ueff_MF} can be seen as a zero-dimensional field theory, and the number of degrees of freedom is reduced from infinity to one \cite{Laine:2016hma}.
In the infinite volume limit (or at $T=0$), the only nonzero weight in \Eq{Eq:Ueff_MF} corresponds to the minimum of $\mathcal{U}_\eff^{}(\bar\phi)$ as the weight $P(\bar\phi)$ reduces to the normalized sum of Dirac delta functions (two at the first order transition, one otherwise). 
This reproduces the usual mean-field model framework, where $\langle\phi\rangle=\bar\phi_\mathrm{min}$ is constant and homogeneous, satisfying the field equation 
$\left( \partial \mathcal{U}_\eff(\phi) / \partial\phi\right)|_{\phi=\bar\phi_\mathrm{min}}=0$.
However, if the system size (and the temperature) is finite, the integral over the mean-field $\bar\phi$ may be performed. An illustration of how the peaks in $P(\phi)$ ``melt'' as $L$ decreases is given in Fig.~\ref{Fig:weights}. 

In most finite-volume studies within effective models at mean-field approximation, the finite-size effects are included primarily via momentum-space constraints (on the fermionic fluctuations). Hence, the expectation value $\langle\phi\rangle$ is fixed to the minimum of the effective potential even at finite sizes, and a second- or first-order phase transition can be present. The inclusion of the zero-mode integration provides a minimal approach that exhibits finite-size rounding and finite-size scaling (albeit within the mean-field universality class). 

To systematically improve the mean-field approximation, one can include fluctuations around the background field perturbatively and integrate out the nonzero modes, which at leading order yields Gaussian functional integrals \cite{Jackiw:1974cv, PhysRevD.9.3320}. This procedure leads to a generally nonlocal effective action, which requires functional methods to evaluate. Restricting the leading-order Gaussian to a constant, homogeneous background field yields a local Gaussian approximation. This is a standard extension of the mean-field approximation \cite{Kovacs:2021kas, Kovacs:2025soz}, as the resulting effective potential is tractable in a similar framework.
However, while the local Gaussian approximation accounts for higher-momentum mesonic modes, it evaluates fluctuations only on a fixed, spatially uniform background $\bar{\phi}$. This defines a constraint effective potential \cite{Fukuda:1974ey, Fukuda:1975ib, ORaifeartaigh:1986axd}, probing the effective action only for homogeneous fields and thus omitting kinetic energy and spatial correlations.

Here, we adopt a different strategy and include a static but spatially varying classical configuration via an ansatz. The inclusion of a specific spatial dependence of the field can be motivated by the solutions obtained in a non-local treatment \cite{Lo:2026xuc}. 
In this work, we only prescribe a spatial profile for the field, without requiring it to solve the equation of motion. 
Specifically, we assume a form $\phi(\vec x)=\bar\phi f(\vec x)$ with a smooth, dimensionless function $f(x)$. Here $\bar\phi$ is an amplitude, related to the zero mode as $\bar\phi=\phi_0/f_0$ by imposing the normalization $\int_{L^d} d^dx f(\vec x)=f_0 V$.
For a finite system, it is convenient to choose $f(x)$ such that it vanishes smoothly on the boundaries and is dominated by the low-momentum modes. Under the assumption that $f(x)$ varies on the scale of the system size $L$, the gradient contribution scales as $\int_{L^d} d^dx (\nabla\phi(\vec x))^2\propto \bar\phi^2 L^{d-2}$.
The integration over $\bar\phi$ may still be performed with an effective action that takes the form for $d=3$
\be \label{Eq:Seff}
S_\eff^\mathrm{} (\bar\phi)= \beta V \left( \frac{c}{2}\, \frac{\bar\phi^2}{L^2}  + \mathcal{U}_\eff^{} (\bar\phi)\right) \,,
\ee 
where $c$ is a constant defined by the choice of $f(x)$. 
While this modification does not alter the asymptotic finite-size scaling behavior at large volumes, the presence of the gradient contribution modifies the finite-size effects at small sizes.

In this work, we mainly use a simple framework based on $\mathcal{U}_\eff^{}$, while qualitatively discussing the effect of having different model choices and that of the gradient term within $S_\eff^{}$.

\subsection{Model} \label{Sec:model}

To perform a numerical calculation, we derive an effective potential that describes the system to be studied.
Since our goal is to discuss the criticality in a chiral phase transition within a simple setup, we use an effective model with one bosonic and one fermionic field that resembles the $N_f=1$ quark-meson model. The Lagrangian reads 
\be \label{Eq:Lag}
\mathcal{L}=\frac{1}{2} \partial_\mu \phi \partial^\mu \phi - V_\cl (\phi) + \bar \psi \left(i \gamma^\mu \partial_\mu - \mu_q \gamma_0 - g \phi \right) \psi \,,
\ee 
with the classical potential
\be 
V_\cl(\phi) = \frac{1}{2} m^2 \phi^2 + \frac{\lambda}{4} \phi^4 - h \phi \,.
\ee 
We consider symmetric quark matter with the quark chemical potential $\mu_q=\mu_B/3$ and use the baryon chemical potential primarily.
By integrating out the fermions, one obtains the usual functional determinant. Then the partition function is in the form of \Eq{Eq:Z_euc_def} with the Euclidean action
\be \label{Eq:S_E}
S_{}[\phi]= \int_x \left(\frac{1}{2} (\partial_\tau \phi)^2 + \frac{1}{2} (\vec{\nabla} \phi)^2 + V_\cl (\phi)\right) - \Tr \ln \mathcal{S}^{-1} [\phi] \,,
\ee 
where $\mathcal{S}^{-1}$ is the inverse fermion propagator. 

When only the zero mode of the mesonic field is kept, the $\int_x$ integration becomes trivial, yielding only a four-volume $\beta V$, and the effective potential can be defined as $ \mathcal{U}_{\eff}^{} (\bar\phi)=S_{}(\bar\phi)/(\beta V)$. In this case, the kinetic contribution is absent. In the fermionic part, we perform the Matsubara sum for the finite temperature calculation. Thus, the effective potential becomes
\be \label{Eq:Ueffqm} \begin{split}
\mathcal{U}_{\eff}^{} 
(\bar \phi,T,\mu_{B},L) =&  V_\cl (\bar\phi) + V_{\bar qq}^v(\bar\phi,L) \\&+ V_{\bar qq}^m(\bar\phi,T,\mu_{B},L)\,,
\end{split} \ee
where the last two terms are the fermionic vacuum and matter part, respectively. These terms are generally size-dependent through the effect of the spatial extent on the momentum space. The fermionic vacuum contribution, given by
\be \label{Eq:Vqqvac}
V_{\bar qq}^v(\bar\phi,T,L) = - 2 \int_{k,L} E(k) \,
\ee 
with $E(k)=\sqrt{k^2+m_q^2}$ and $m_q=g\bar\phi$, is UV divergent and needs to be regularized.
The fermionic matter part reads
\be \label{Eq:Vqqmat}
V_{\bar qq}^m(\bar\phi,T,\mu_{B},L) = - 2 T N_c \int_{k,L} \left(\ln g^+ + \ln g^- \right) \,,
\ee 
where $g^\pm = (1+e^{-\beta E^\pm (k)})$ with $E^\pm(k)=E(k)\mp\mu_q$.
Here we denote the momentum integral (or sum) by $\int_{k,L}$, which reduces to $\int_{k,L\to\infty}=\int \frac{d^3k}{(2\pi)^3}$ for $L\to\infty$. At finite sizes, $\int_{k,L}$ is a summation over the modes determined by the spatial boundary conditions or, in a simplified scenario, an integration with a low-momentum cutoff. These cases and their treatment in a mean-field model are discussed in detail in \cite{Kovacs:2023kbv}.

In our main approach, we neglect the momentum-space constraints and hence the intrinsic size dependence of the potential to isolate the role of the zero mode integration. Moreover, we keep only the classical potential and the fermionic matter part. 
The simplified effective potential reads
\be \label{Eq:Ueff_infty}
\mathcal{U}_\eff^\infty (\bar\phi,T,\mu_{B}) = V_\cl (\bar\phi) + V_{\bar qq}^m(\bar\phi,T,\mu_{B},L\to\infty)\,.
\ee
Consequently, the only volume dependence of this model is in the prefactor of $\mathcal{U_\eff^\infty}$ in $\mathcal{Z}$ in \Eq{Eq:Ueff_MF} and hence in the free energy.
While the fermionic vacuum contribution is absent in \Eq{Eq:Ueff_infty}, its presence yields different qualitative behavior only when the momentum is discretized. This approximation provides a minimal framework to study finite-size effects from the statistical weighting, which is the main focus of the present work.

To incorporate the effect of the momentum space constraints, we consider the effective potential
\be  \label{Eq:Ueff_L_dep} \begin{split}
\mathcal{U}_\eff^{L} (\bar\phi,T,\mu_{B},L)=&V_\cl (\bar\phi) + \Delta_L V_{\bar qq}^v(\bar\phi,L) \\&+ V_{\bar qq}^m(\bar\phi,T,\mu_B,L)\,,
\end{split} \ee 
which involves the size-dependent fermionic matter fluctuations and the finite-size correction of the fermionic vacuum part. The latter is defined by 
\be \label{Eq:VvacLcorr} \begin{split}
\Delta_L V_{\bar qq}^{v}(\bar\phi,L) = -2 \Bigg(&\frac{1}{L^3}{\sum_{\vec{n}}}^\Lambda E_{\vec{n}} - {\int}^\Lambda \frac{d^3k}{(2\pi)^3}  E(\vec{k})\\& - \frac{1}{L^3}{\sum_{\vec{n}}}^\Lambda |\vec{p}| + {\int}^\Lambda \frac{d^3k}{(2\pi)^3} |\vec{p}|\Bigg)\,,
\end{split} \ee 
which is convergent for $\Lambda\to\infty$, hence can be determined with a required precision using a sufficiently large cutoff. By including only the $\Delta_L V_{\bar qq}^{v}$ size-dependent part of the vacuum term, we avoid the $\propto -\bar\phi^4\ln \bar\phi^2$ contribution, which would make the potential unbounded from below at $\bar\phi\to\infty$ and hence the $\bar\phi$ integration not well-defined \cite{Kovacs:2023kbv}. In the infinite volume limit \Eq{Eq:Ueff_L_dep} reproduces \Eq{Eq:Ueff_infty}.
In this work, we only consider the case of periodic boundary conditions (PBC) giving rise to $k_i=2\pi n_i/L$ for $n_i\in\mathbb{Z}$ and $i=1,2,3$.

As we focus on a qualitative investigation of the phase transition, we set the numerical value of the model parameters such that the resulting phase diagram is sensible and features a critical endpoint. To see how the location of the CEP influences the results obtained, we include four sets of parameters 
with fixed $m=-0.155~\text{GeV}^2$ and $h=0.0018~\text{GeV}^3$, but varying $\lambda$ and $g$ as listed in Table~\ref{tab:param_sets}. 
\begin{table}[]
    \centering
    \setlength\extrarowheight{2pt}
    \begin{tabular}{l |c c c c}
         & ~~set A~~ & ~~set B~~ & ~~set C~~ & ~~set D~~ \\
         \hline 
         \hline
        $\lambda$ & 27 & 28 & 29 & 32 \\
        $g$ & 4.60 & 4.60 & 4.55 & 4.50 \\
        \hline
        $T_\mathrm{pc}^{\mu=0}$ & 158.4 & 158.0 & 158.5 & 158.5 \\
        \hline
        $T^\mathrm{CEP}$ & 137.4 & 131.3 & 121.5 & 103.8 \\
        $\mu_{{B}}^\mathrm{CEP}$ & 446.1 & 498.3 & 579.9 & 687.3 
    \end{tabular}    
    \caption{
    The parameter sets used in the numerical calculations. The respective pseudocritical temperatures at $\mu_{B}=0$ and the location of the CEP at $L\to\infty$ (each in units of MeV) are also shown.}
    \label{tab:param_sets}
\end{table}
At $L\to\infty$, each parametrization exhibits almost equal $T_\pc$ at $\mu_{B}=0$ and similar phase boundaries with a critical point. 
If not specified, the parameter set C is used. 

\subsection{Gradient term} \label{Sec:gradient}

In principle, the gradient term of the scalar field generates nontrivial contributions to the one-loop effective action~\cite{PhysRevD.9.3320, Lo:2026xuc} for the background field. 
To incorporate it, we adopt an ansatz for a static but nonhomogeneous classical field $\phi(\vec x)=\bar\phi f(\vec x) $. For definiteness, we chose a specific form $f_{\cos} (\vec x)=f_0 \prod_{i} (1+\cos(\frac{2\pi x_i}{L}))$ giving rise to
\be \label{Eq:Skin}
S_\mathrm{kin}(\bar\phi)= \beta \frac{\bar\phi^2}{2} \int d^3x (\vec \nabla f_{\cos}(\vec x))^2 = \beta L\, \frac{c\bar\phi^2}{2} \,
\ee 
with $c=f_027\pi^2/2$. The function $f_{\cos}(\vec{x})$ is chosen such that the field vanishes at the boundaries of the $[-L/2,L/2]^3$ finite volume, while its overall normalization is absorbed into the coefficient $c$, discussed below. With this ansatz we extend our approach based on Eq.~\eqref{Eq:Ueff_infty} to the form in Eq.~\eqref{Eq:Seff}
\be \label{Eq:Sgrad}
S_\eff^\mathrm{grad}(\bar{\phi},T,\mu_B,L) = S_\mathrm{kin}(\bar{\phi},L) + \beta V \mathcal{U}_\eff^{\infty} 
(\bar \phi,T,\mu_B) \,.
\ee
The pure mean-field case and \Eq{Eq:Ueff_infty} can be recovered by having $f_\MF(\vec x)=1$. Using $f_{\cos} (\vec x)$ in the classical potential raises further numerical factors, which, however, can be merged into the coupling constants when carrying out the parametrization. We assume that the fermionic field couples only to the global mode of the mesonic field to avoid additional structure in the fermionic fluctuations. Since $\phi(x)$ is given by an ansatz, without solving the equation of motion, the relative weight of the gradient term and the effective potential is not yet determined. We set $c$ such that the effective quadratic coefficient (including the gradient term, the quadratic term, and the fermion fluctuations) is negative, hence the minimum with spontaneous symmetry breaking is not absent above $L\approx 1$ fm. Particularly, we investigate $c\leq6.0$ using the parameter set C. While the choice of $c$ affects the sensitivity of finite-size effects to the gradient term, the qualitative behavior is robust and does not depend on its precise value within the same order of magnitude.

\medskip
\medskip
\subsection{Scaling behavior} \label{Sec:scaling}

We introduce the Ising thermal and external field scaling fields, which we denote by $\tau$ and $\eta$, respectively, to avoid confusion with other quantities. From self-similarity in the vicinity of a critical point, the scaling ansatz for the singular part of the free energy density in a system of finite size $L$ is given by
\be \label{Eq:scaling_ansatz}
f_s(\tau,\eta) = L^{-d} \mathcal{F}(\tau L^{y_\tau}, \eta L^{y_\eta})\,,
\ee 
with $y_\tau,y_\eta>0$.
From \Eq{Eq:scaling_ansatz}, the finite-size scaling for the derivatives of the free energy can be derived as   
\be \label{Eq:chietak_def}
\chi^\eta_k = \partial_\eta^k f_s = L^{k\, y_\eta-d} \mathcal{F}_{\eta,k}(\tau L^{y_\tau}, \eta L^{y_\eta})\,,
\ee 
and
\be \label{Eq:chitauk_def}
\chi^\tau_k = \partial_\tau^k f_s = L^{k\, y_\tau-d} \mathcal{F}_{\tau,k}(\tau L^{y_\tau}, \eta L^{y_\eta})\,.
\ee 
The $y_\eta$ and $y_\tau$ scaling dimensions can be determined non-perturbatively using functional renormalization group techniques. By exploiting hyperscaling relations, these are related to the standard critical exponents via $y_\tau=1/\nu$ and $y_\eta=\beta\delta/\nu$.
Therefore, the finite-size scaling, e.g., for the singular part of Ising-like magnetization, is $M=\chi_1^\eta\propto L^{-\beta/\nu}$ while for the susceptibility $\chi_2^\eta\propto L^{\gamma/\nu}$. 
However, in the mean-field approximation in $d=3$ spatial dimensions---which we use in this work---hyperscaling is violated, as can be seen, e.g., from $2-\alpha \neq\nu d$ with the mean-field values of the exponents. Such a scenario is known in systems above the upper critical dimension $d_c$ \cite{Binder:1985_5DIsing, Binder:1985xkp, Privman:1983}. To keep the well-known notation, following Ref.~\cite{Binder:1985_5DIsing}, an effective exponent $\tilde\nu=(\gamma+2\beta)/d=2/3$ can be introduced. This works in the standard scaling formulas expressed in terms of critical exponents, also in the case below the upper critical dimensions, but with mean-field approximation.

The above discussion of the scaling behavior relies on the singular part of the free energy and the derived quantities. However, our free energy in \Eq{Eq:freeenergy_def} also contains a regular part $\Phi_\mathrm{reg}$ that generally scales as $L^d$ and hence, analogously to \Eq{Eq:scaling_ansatz}, $f_\mathrm{reg}(\tau,\eta)\propto L^0$, which is also inherited by the derivatives. Therefore, this contribution dominates the scaling of the free energy itself and of its first derivatives at large sizes, since $y_\eta, y_\tau<d$ in case of the mean-field exponents. 
To study the scaling of these quantities, the regular part shall be subtracted. 

To apply the finite-size scaling ansatz, the parameter space of our model, spanned by $T$, $\mu_B$, and $h$, has to be mapped into the Ising scaling directions. Usually, the external field $h$ (corresponding to the current quark mass) is fixed, and hence, the irrelevant direction is absent on the remaining surface. Then the Ising temperature and Ising external field-like directions can be obtained in the vicinity of the CEP via a linear map in $T$ and $\mu_B$ \cite{Parotto:2018pwx}. However, the scaling implies that the $\eta$ direction has a nonzero overlap with each $\mu_B$, $T$, and $h$. Each of these exhibits an external field-like scaling, and the generalized susceptibilities, $\chi_k$, $\chi^B_k$, and $\chi_k^T$ (defined by the $k^\mathrm{th}$ derivative in the $h$, $\mu_B$, and $T$ directions, respectively) share the same volume dependence. Consequently, the order parameter, chiral susceptibility, baryon number, entropy, and energy density (and their fluctuations) show the scaling dominated by the $\eta$-derivatives. It is also expected that the temperature-like scaling will be observable only in directions along the phase boundary (or phase boundary surface if $h$ is considered) in a narrow window for a linear sigma model \cite{Kovacs:2006ym}. Similarly, the trivial scaling is recovered only when probed along the irrelevant or marginal direction $u$ tangent to the critical line.
While the universal critical exponents are invariant under the specific choice of the $T,\mu_B,h-\tau,\eta$ mapping, the non-universal scaling functions depend on the trajectory in the parameter space.

\section{Finite size dependence} \label{sec:results}

Using the effective model outlined in \Sec{sec:method}, we study the size dependence of the phase diagram, different physical quantities, and their fluctuations. We start with the chiral condensate and susceptibilities and discuss their finite-size scaling.
We note that, as discussed in the previous section, the baryon number fluctuations show the same scaling properties.
To ease the notation, we will omit the bar notation and use $\phi$ for the constant mode of the field over which the integration is performed.

\subsection{Chiral condensate and susceptibility} \label{Sec:chiral}

\begin{figure}
    \centering
    \includegraphics[width=0.95\linewidth]{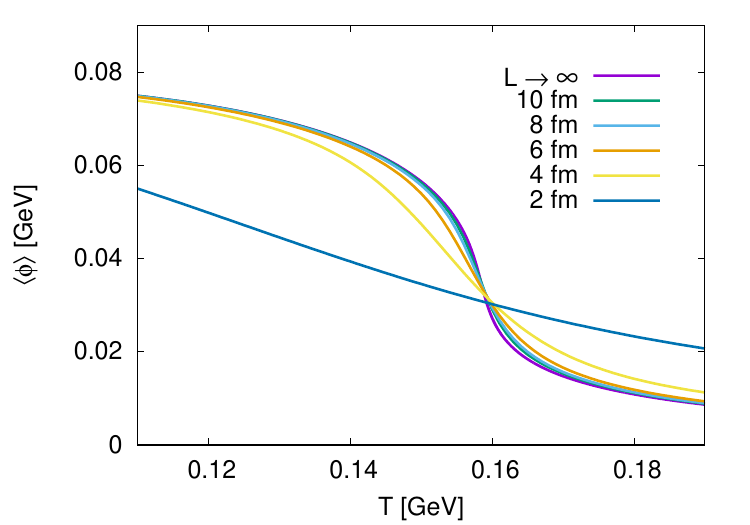}
    \caption{The chiral condensate $\langle\phi\rangle$ at $\mu=0$ for different system sizes and in the thermodynamic limit using $\mathcal{U}_\eff^\infty(\phi)$.}
    \label{Fig:phi_Lfin_mu0}
\end{figure}

First, we discuss the general effect of the finite size on the chiral condensate purely from the nonperturbative integral over the mean-field using $\mathcal{U}_\eff^\infty$, which immediately highlights the limitations and advantages of this approach.
At non-vanishing temperatures, it is clear in our formalism that the weights in \Eq{Eq:phi_avg_formal} tend to become equal for any $\phi\in(-\infty,\infty)$ for $L\to0$ since the exponential tends to unity, independently of the field. Therefore, in the chirally broken region, $\langle\phi\rangle$ decreases for a small, decreasing $L$.\footnote{In the chirally restored region close to $T_\pc$, due to the emerging nonzero weight of the $\phi>\phi_0\sim0$ contribution, $\langle\phi\rangle$ increases unless the size is very close to zero.} As shown in \Fig{Fig:phi_Lfin_mu0}
for $\mu_{B}=0$, this smoothens the transition and also gives rise to a decrease of the pseudocritical temperature as the crossover between the phases shifts for $L\lesssim 10$ fm. However, in $T\to 0$ there are no finite-size effects due to the statistical weighting, since the full four-volume $\beta V$ tends to infinity. Therefore, the vacuum value of $\langle\phi\rangle$ is constant in $L$. This leads to the main caveat compared to lattice results, where an exponential rise of the chiral condensate is found for very small sizes \cite{RubenKara:2024krv, Borsanyi:2025lim}. The discrepancy reflects the limitation of using the infinite-volume effective potential $\mathcal{U}_\eff^\infty$. The normalized condensates already show the same dependence at finite $T$ and $L$, which anticipates that the phase boundary behaves similarly when the size decreases. However, the size dependence of $\langle\phi\rangle$ also determines that of the chiral susceptibility $\chi$, for which we see a decreasing maximum due to the softening of the transition, unlike on the lattice. 

Interestingly, the ChPT calculations provide a decreasing trend in the quark condensate \cite{Gasser:1986vb, Bijnens:2006ve}, even in the regime of the $\epsilon$ expansion \cite{Gasser:1987ah, Damgaard:2008zs},
which becomes relevant for $L\lesssim2$~fm for the physical pion mass.
One can see an exponential dependence in ChPT---similar to that in the lattice calculations, but with opposite sign---with also an interesting interplay with a finite magnetic field \cite{Adhikari:2023fdl}. Functional methods with momentum-space constraints also predict the decrease of the chiral condensate for very small sizes \cite{Luecker:2009bs}.

We emphasize that these effects in the crossover region are not due to the scaling, but really are finite-size effects. In the model with $\mathcal{U}_\eff^\infty$, they arise purely from the weighting in the nonperturbative integral. 

\begin{figure}
    \centering
    \includegraphics[width=0.95\linewidth]{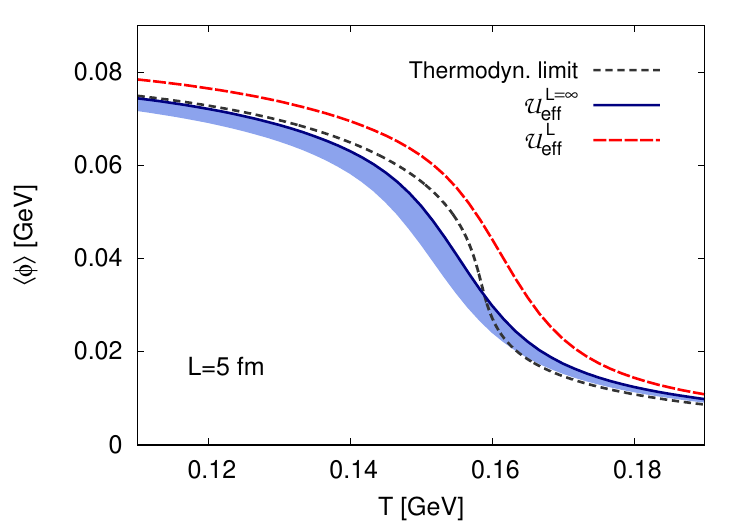}
    \caption{The effect of the modeled gradient term for $c\leq6$ (represented by the shaded area) and the discretization of fermion fluctuations on the chiral condensate $\langle\phi\rangle$ at $\mu_B=0$ and $L=5$ fm. The result in the thermodynamic limit is shown for comparison.}
    \label{Fig:phi_Lfin_mod_mu0}
\end{figure}

When the momentum-space constraints are included, the effective potential itself becomes $L$-dependent. However, its effect may depend on the chosen type of constraint, and the treatment of the vacuum contribution \cite{Kovacs:2023kcn, Wang:2018qyq}.
Using $\mathcal{U}_\eff^{L}$ with PBC, the value of $\langle\phi\rangle$ at low temperatures increases for decreasing $L\lesssim 10$ fm. This can be seen in \Fig{Fig:phi_Lfin_mod_mu0}, where we also show the finite-size result with no discretization and the result in the thermodynamic limit for comparison. Due to the increasing $\langle\phi\rangle_{T=0}$, the transition becomes steeper for the decreasing size, hence giving rise to an increasing peak of the chiral susceptibility. In our model, this effect is independent of the chosen boundary conditions for the fermions (similar as in Ref.~\cite{Kovacs:2023kcn}) for $L\gtrsim 1$ fm. This can be understood, since the finite-size correction in \Eq{Eq:VvacLcorr} is governed by the nonzero low-momentum modes, contrary to calculations in the Nambu--Jona-Lasinio model \cite{Wang:2018qyq}, where the regularization may partially suppress these modes. Our findings are generally in line with the lattice results, except for the increasing pseudocritical temperature. In this aspect, the size dependence of $\langle\phi\rangle$ with momentum discretization is similar to the problem of inverse magnetic catalysis. There, the mean-field framework cannot describe the increase of the condensate at low temperatures and its decrease at high temperatures for increasing magnetic field simultaneously \cite{Bruckmann:2013oba}. This problem can be solved in the case of a magnetic field by applying a self-consistent treatment \cite{Lo:2021buz}. Whether such an approach resolves the discrepancy for momentum discretization would require further investigation.

To investigate the effect of the modeled gradient contribution, we employ the ansatz introduced in Eqs.~\eqref{Eq:Skin}-\eqref{Eq:Sgrad}. 
Since the gradient term compensates the quadratic term in the potential, it effectively reduces $\langle\phi\rangle$ as the system size decreases. The pseudocritical temperature also decreases substantially, while the transition becomes softer, as shown in \Fig{Fig:phi_Lfin_mod_mu0} with the shaded region.

\subsection{Finite size scaling}
 
\begin{figure}
    \centering
    \includegraphics[width=0.95\linewidth]{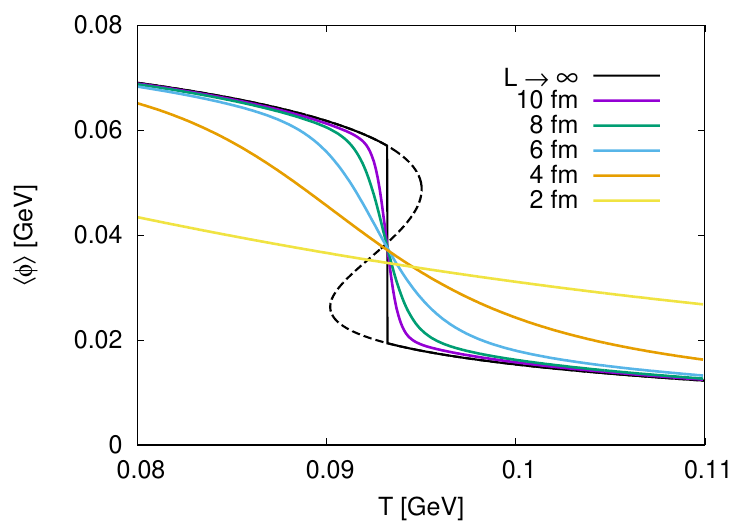}
    \caption{The chiral condensate $\langle\phi\rangle$ at a first-order transition for different system sizes and in the thermodynamic limit using $\mathcal{U}_\eff^\infty(\phi)$. The dashed curve shows the $L\to\infty$ meta- and unstable solutions. }
    \label{Fig:phi_Lfin}
\end{figure}

At high chemical potentials, the phase transition for $L\to\infty$ is of first order in our model. However, at finite sizes, there is always only a single solution and the transition is smooth for any finite $L$, as shown in \Fig{Fig:phi_Lfin}.\footnote{Due to the discussed unchanged behavior at $T=0$, the phase transition at vanishing temperature remains sharp and gives an $L^d/T$ divergence \cite{Binder:1992pz}.} 
A similar rounding of the transition was also recently discussed in Ref.~\cite{An:2025kaw} within a different framework.

\begin{figure}
    \centering
    \includegraphics[width=0.95\linewidth]{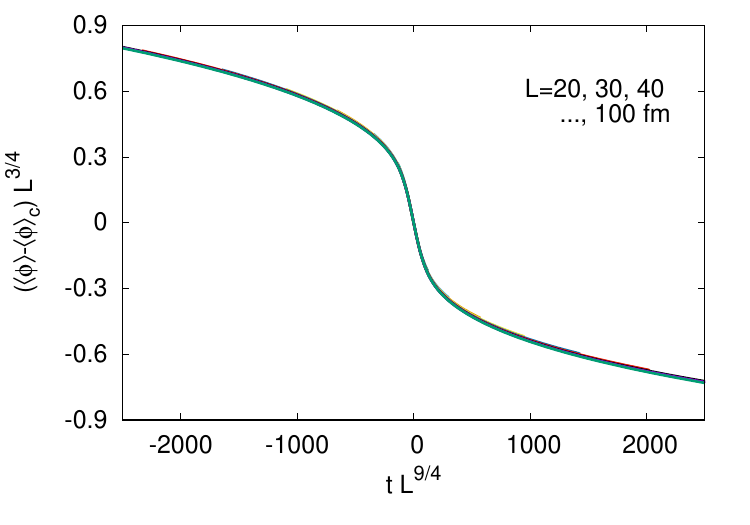}
    \includegraphics[width=0.95\linewidth]{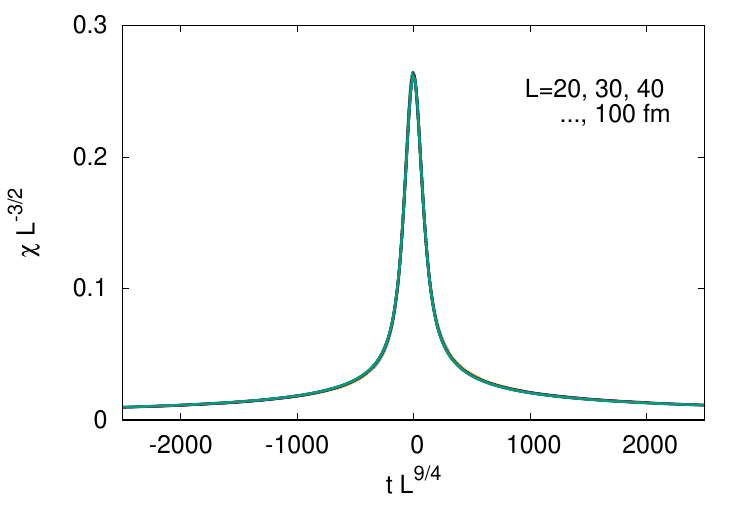}
    \caption{The $L=30,\,32,\,34,\ldots,\,100$~fm results collapsing to a single curve for the singular part of the chiral condensate (top) and the chiral susceptibility (bottom) as a function of the scaled subtracted temperature. The $L=10$ and $20$~fm curves are presented to illustrate the onset of deviation from the scaling regime.}
    \label{Fig:scaling_CEP}
\end{figure}

Since the phase transition becomes smooth in our model, the finite-size effects diminish the susceptibility at the critical point, and we can investigate the scaling behavior. As the momentum discretization and the modeled gradient term do not alter the scaling behavior, and their effect is already negligible in the corresponding size range, here we use the simplest scenario with $\mathcal{U}_\eff^\infty$. As discussed before, in the case of the chiral condensate, one shall take care of the regular part of the free energy. It is sufficient to approximate $f_\mathrm{reg}$ with a linear ansatz,\footnote{One may fit $f_\mathrm{reg}+f_s$ in the infinite volume case. In the $T,\mu,h$ parameter range of interest, the linear ansatz works properly. Higher order terms give only negligible corrections in $\Phi,~\langle\phi\rangle$, and $\chi$, while terms beyond quadratic order cannot even be fitted reliably due to their vanishingly small contribution.} which brings only a constant $\langle\phi\rangle_c\equiv\langle\phi\rangle|_{T^\CEP,\mu^\CEP}$ subtraction from $\langle\phi\rangle$, while $\chi$---which is anyway dominated by the singular part---remains unchanged. 
The obtained finite-size scaling is presented in \Fig{Fig:scaling_CEP} through the $L\to\infty$ critical point as a function of the reduced temperature $t=(T-T_c)/T_c$ scaled with the appropriate power of $L$. As discussed in \Sec{Sec:scaling}, since both $h$ and $T$ have overlap with the $\eta$ direction, we need to use the external field-like scaling exponents. The results collapse to a single curve, which shows scaling for $L\gtrsim  20~\text{fm}$. For these sizes, the shift in the transition temperature is negligible, while for $L\lesssim 20$~fm, the finite-size effects emerge.

\begin{figure}
    \centering
    \includegraphics[width=0.95\linewidth]{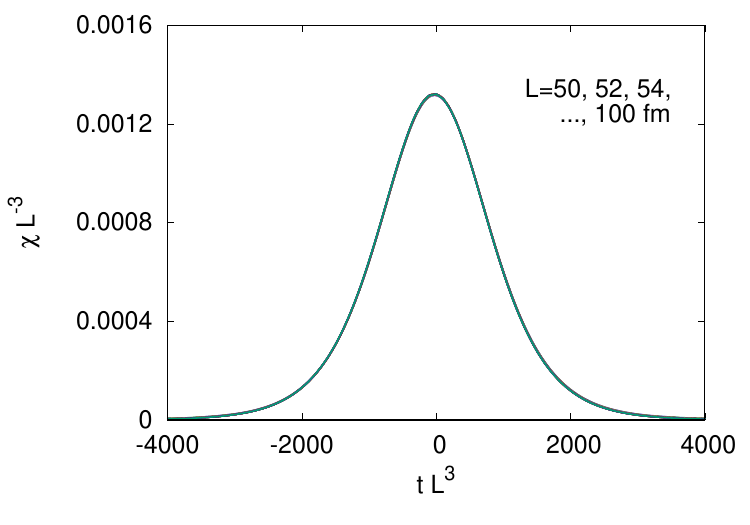}
    \caption{The scaled chiral susceptibility as a function of the scaled temperature in the neighborhood of the first-order transition.}
    \label{Fig:scaling_1st}
\end{figure}

It is interesting to study the susceptibility also in the vicinity of the $L\to\infty$ first-order transition. It turns out that the coexistence---in our case, the presence of two degenerate minima of the potential---gives rise to a nontrivial volume dependence. While the chiral condensate remains $\langle\phi\rangle\propto L^0$, the susceptibility obtains a $\chi\propto L^d$ (generally $\chi_k\propto L^{k\,d}$) scaling in the vicinity of $T_c$. 
This scaling behavior is shown in \Fig{Fig:scaling_1st}. 
The phenomenon can be easily understood by using a double Gaussian approximation for the probability distribution \cite{Binder:1985xkp}, which we also review in Sec.~\ref{sec:doubleG}. Due to the $\chi_k\propto L^{k\,d}$ size-dependence driven by the coexistence in the first-order case, the peak of the chiral susceptibility monotonously increases as $\mu_B$ increases along the phase boundary. Thus, no ``apparent critical endpoint'' can be identified merely as a maximum of $\chi$ as done in functional methods with momentum-space constraints \cite{Almasi:2016zqf, Tripolt:2013zfa, Bernhardt:2021iql}. To locate the $L\to\infty$ critical endpoint, one needs a different strategy, such as using the Binder cumulant, to be discussed in detail in \Sec{Sec:along_pb}. 

The critical mode at the chiral CEP (in the absence of further condensates \cite{Haensch:2023sig}) corresponds to the scalar singlet $\sigma$, here $\phi$ itself. The mass squared of this field can be identified with the inverse of the chiral susceptibility. It is usual to utilize this relation to define the meson mass at finite system sizes by $m_\phi^2=\chi_2^{-1}$ (or vice versa), which is nonzero when $\chi_2$ does not diverge. At finite sizes, this is not equivalent to the expectation value of the curvature of the effective potential. The mass, defined with the susceptibility, does not have a minimum along the phase boundary around the $L\to \infty$ critical point for finite $L$. Since the expectation value of the curvature of $\mathcal{U}_\eff$ has no such effect from the coexistence, it exhibits a minimum. However, this quantity does not correspond to the two-point function, from which the curvature mass might be obtained in the vanishing momentum limit.

\medskip
\medskip
\subsection{Effects of the coexistence} \label{sec:doubleG}

At a first-order transition in finite volume, the susceptibility diverges as $\chi\propto L^d$ in the limit $L\to\infty$. This divergence originates from the coexistence~\cite{Binder:1992pz}, which we discuss in a double Gaussian approximation \cite{Binder:1984llk, Binder:1992pz}. To employ its results in our approach, we also provide a definition for $\eta{(T,\mu_B,h)}$ that applies at any point on the phase boundary.

To outline the double Gaussian approximation, we start with a general potential $\mathcal{U}(\phi)$ that exhibits minima at $\phi=\sigma_a$ and $\sigma_b$ with ``susceptibility'' (inverse of the curvature) $\chi_a$ and $\chi_b$, respectively, at the first-order transition. By expanding $\mathcal{U}$ at $\eta=0$ in $\phi$ around its minima, one can write the probability density in the form\footnote{The simplified form, e.g., shown in Ref.~\cite{Binder:1992pz}, can be written in the symmetric $\chi_{a}=\chi_{b}$ case.}
\be \label{Eq:P_DG} \begin{split}
P_{DG}(\phi)=&\frac{A}{(2 \pi \Tau)^{1/2}} \\&\times\Big[ e^{(\eta\,\sigma_a+\frac{1}{2}\eta^2\chi_a)\frac{L^d}{\Tau}}e^{-\frac{(\phi-\sigma_a-\eta \chi_a )^2}{2 \chi_a}\frac{ L^d}{\Tau}}\\&\quad~ +e^{(\eta\,\sigma_b+\frac{1}{2}\eta^2\chi_b)\frac{L^d}{\Tau}}e^{-\frac{(\phi-\sigma_b-\eta \chi_b )^2}{2 \chi_b}\frac{ L^d}{\Tau}}\Big]\,,
\end{split} \ee
with normalization constant $A$ being
\be \begin{split}
A=L^{d/2}&\Big[\chi_a^{1/2}e^{(\eta\,\sigma_a+\frac{1}{2}\eta^2\chi_a)L^d/\Tau}\\&+\chi_b^{1/2}e^{(\eta\,\sigma_b+\frac{1}{2}\eta^2\chi_b)L^d /\Tau}\Big]^{-1}
\end{split} \ee
to satisfy $\int_{-\infty}^{\infty}d\phi \,P_{DG}=1$. $\Tau$ is the (not reduced) Ising temperature.
For sufficiently large sizes, $P_{DG}$ is expected to be a good approximation of the full probability distribution, since the integration is dominated by the minima.
Using $P_{DG}(\phi)$, one can calculate
\be \label{Eq:DGphi}
\langle\phi\rangle_{DG} = \lambda_a U_a + \lambda_b U_b
\ee
and
\be \label{Eq:DGphi2}
\langle\phi^2\rangle_{DG} = \frac{\Tau}{L^d} \left( \chi_a U_a + \chi_b U_b \right) + \lambda_a^2 U_a + \lambda_b^2 U_b
\ee\\
by evaluating the Gaussian integrals over $\phi$. Here $\lambda_i=\sigma_i+\eta\,\chi_i$, $U_i=W_i/(W_i+W_{j})$ and $W_i=\sqrt{\chi_i}\exp\lbrace(\eta\,\sigma_i +\eta^2\chi_i/2)L^d/\Tau\rbrace$ for $i,j=a,b$ with $j\neq i$. In \Eq{Eq:DGphi} it can be seen that $\langle\phi\rangle_{DG}$ is a weighted sum of the contribution from the two minima, depending not only on $\sigma_i$, but also on $\chi_i$. At the transition, i.e., $\eta=0$ the weights are exactly $1/2$ if $\chi_a=\chi_b$. 

Using $\chi_{DG}=L^d/\Tau (\langle \phi^2\rangle_{DG}-\langle \phi\rangle_{DG}^2)$ one obtains for the chiral susceptibility
\be \label{Eq:DGchi} \begin{split}
\chi_{DG}=& \chi_a U_a + \chi_b U_b + \frac{L^d}{\Tau}(\lambda_b - \lambda_a)^2 U_aU_b\,.
\end{split} \ee
The first terms come from the contribution in \Eq{Eq:DGphi2} where the four-volume factor cancels out. This gives simply the weighted sum of the susceptibilities in the two minima, similar to the structure of  $\langle\phi\rangle_{DG}$. The $\propto L^d$ part arises from the cross products of the terms describing the separate minima and
characterizes fluctuations between them.
Along the phase boundary ($\eta=0$, hence $\lambda_i\to\sigma_i$) this contribution depends on the difference $(\sigma_b-\sigma_a)^2$. Therefore, when approaching the critical endpoint with increasing $\Tau$, the effect of coexistence clearly becomes absent at the CEP. The weight $U_i$ of the local minimum becomes exponentially suppressed for large volumes, and the width of the peak of $\chi_{DG}(\eta)$ scales as $L^{-d}$.

To make use of the double Gaussian formulas in our approach, we need to map the $T$, $\mu_{B}$, and $h$ variables to $\Tau$ and $\eta$. We want to define a mapping that can be used at any point of the phase boundary surface and fits the construction of the double Gaussian form in \Eq{Eq:P_DG}.
We start with a general effective potential in the form 
\be \label{Eq:formofU}
\mathcal{U}(\phi;T,\mu_{B},h)=u(\phi ;T,\mu_{B})-h\phi\,,
\ee
where the linear field dependence is separated, and $u$ is assumed to be an even function of $\phi$. Here $h$ has an analogous role to the Ising external field for the transition at $h=0$ (the order of which depends on the value of $T$ and $\mu_{B}$).
We aim to separate a similar linear term on the phase boundary associated with a chiral transition to identify the corresponding external-field direction from its coupling. This transition has a nontrivial $T,\mu_{B},h$ surface, with $T_c$, $\mu_{B,c}$, and $h_c$ each possibly being nonzero (parameters with subscript $c$ mark here a given point on the phase boundary surface). First, we can define $\tilde{\phi}=\tilde\phi(T,\mu,h)$ such that $\mathcal{U}$ has an extremum at $\tilde \phi$ along the phase boundary surface (maximum at first-order, minimum at crossover, and flat minimum at second-order transition), hence $\mathcal{U}^\prime(\tilde\phi,T_c,\mu_{B,c},h_c)=u^\prime (\tilde \phi,T_c,\mu_{B,c})-h_c=0$ with the prime denoting the field derivative and $f^\prime (x_c)\equiv(\partial_\phi f(x))|_{x_c}$. 
To find the coupling of the linear field dependence, one can expand the potential in $\phi$ around $\tilde\phi$ or, equivalently, in the shifted field $\varphi=\phi-\tilde\phi$ around zero to get
\be \label{Eq:htoetainU}\begin{split}
\mathcal{U}(\phi;T,\mu_{B},h)=& \mathcal{U}(\tilde\phi;T,\mu_{B},h)+\mathcal{U}^\prime(\tilde\phi;T,\mu_{B},h) (\phi-\tilde\phi) \\&+ \frac{1}{2} \mathcal{U}^{\prime\prime}(\tilde\phi;T,\mu_{B},h) (\phi-\tilde\phi)^2 + \ldots \\
=&\tilde{\mathcal{U}}(\varphi;T,\mu_{B},h) - \eta(T,\mu_{B},h)\varphi\,.
\end{split}\ee 
Here we implicitly defined $\tilde{\mathcal{U}}$ and 
\be \label{Eq:defeta}
\eta(T,\mu_{B},h)\equiv-\mathcal{U}^\prime(\tilde \phi;T,\mu_{B},h)=h-u^\prime (\tilde \phi;T,\mu_{B})\,,
\ee 
and also simplified the parameter-dependence through $\tilde\phi=\tilde\phi(T,\mu_{B},h)$ to ease the notation. 
The expression in \Eq{Eq:defeta} can provide a definition for an ``external~field'' $\eta$ along the whole phase boundary surface, although with the caveat that the potential is not fully symmetric in $\varphi$ around $\varphi=0$. 
We emphasize, however, that the effects of this asymmetry 
(other than the properties of the minima, such as their curvature) 
are suppressed for sufficiently large $L$, which justifies our strategy.

\begin{figure}[]
    \centering
    \includegraphics[width=0.95\linewidth]{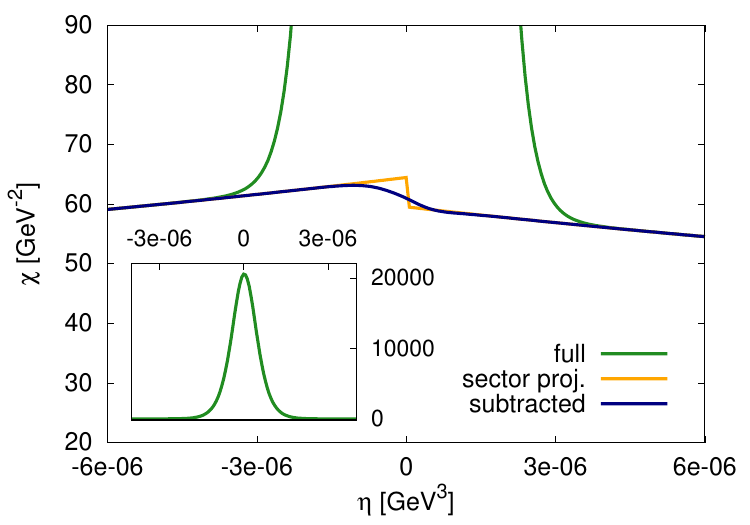}
    \caption{The chiral susceptibility as a function of $\eta(T,\mu_{B},h)$ across the first-order transition. The green line is the full result calculated at $L=250$ GeV$^{-1}\approx 50$ fm (also shown in the inset), while the blue curve is the remnant after subtracting the divergent part as discussed in the text.
    The orange line is the sector projected result for the two minima.}
    \label{Fig:coex_removed_peak}
\end{figure}

To determine the external field-like direction, we need to relate the change of $\eta(T,\mu_{B},h)$ defined in \Eq{Eq:defeta} to the change of its arguments. The simplest approximation is to compare $\partial_x\eta$ for $x=T$, $\mu_{B}$, and $h$, which is equivalent to expanding $\eta$ in $T$ and $\mu_{B}$ up to linear order. 
This gives, for instance, $\partial\eta/\partial T\approx-0.078~\text{GeV}^2$, $\partial\eta/\partial\mu_{B} \approx-0.013~\text{GeV}^2$, and $\partial\eta/\partial h=1$ at the first-order transition at $T_c=0.1094$ GeV, $\mu_{{B},c}=0.66$ GeV, $h_c=0.0018$ GeV$^3$, which we use below.\footnote{From such an approximation evaluated at the CEP, it is clear again that $\eta$ has a nonzero overlap with each $x=T$, $\mu_{B}$, and $h$. Note that the magnitude of the $d\eta/dx$ ratios shall not be directly compared due to the different dimensionality and absolute magnitude of the parameters themselves.}
To turn to $\chi(T)$, presented in the previous section, would require relating also $T$ to $\eta$ and $\Tau$.
It is more convenient and straightforward to discuss $\chi(\eta)$ directly, using $\eta(T,\mu_{B},h)$ evaluated above. We note that, in both cases, the value of $\Tau$ still has to be determined. We obtain this by simply fitting $\chi_{DG}(\eta)$ to the calculated $\chi_\text{calc}(\eta)$, which yields $\Tau\approx0.11$~GeV.\footnote{The subtracted temperature $\tau$ might be approximated analogously as in a Landau-type potential. Then, it is related to the second field derivative of the potential $u^{\prime\prime} (\tilde \phi,T,\mu_{B})$.} We define the subtracted susceptibility by 
\be 
\chi_\text{subt}=\chi_\text{calc} - \frac{L^d}{\Tau}(\lambda_b - \lambda_a)^2 U_aU_b \,.
\ee 
where $\chi_\text{calc}$ is the numerical result along the parametrized $\eta(T,\mu_{B},h)$ direction. The second term is calculated using the fitted $\Tau$, the location of the minima $\sigma_{a,b}$ and the respective curvatures $\chi_{a,b}$ at the transition, and $\eta(T,\mu_{B},h)$ as above.
While $\chi_\text{subt}\propto L^0$ is thus obtained, it can still show a fluctuation in the vicinity of the first-order transition depending on the precision of $\eta$ and of the fit. By carefully fine-tuning the $\eta$ direction and precisely pinpointing the CEP, this fluctuation can be suppressed to a rather smooth curve as shown in \Fig{Fig:coex_removed_peak}. The necessity of the very precise fitting and the remaining fluctuations can be understood to originate from approximations made.
Namely, the potential is not symmetric in $\varphi$ around $\varphi=0$, while $\eta$ and $\Tau$ depends nonlinearly on $T$, $\mu_{B}$, and $h$. Also, unless the volume is extremely large, there is a correction to the double Gaussian approximation itself. To sample the contribution from the two minima separately, one can use the sector projection by integrating separately for $\phi\leq\tilde{\phi}$ and $\phi>\tilde{\phi}$. For sufficiently large sizes, this can reproduce the contribution from the two (even local) minima separately, between which $\chi_\text{subt}$ interpolates. From the other perspective, it is also clear that the sector projected result always misses the contribution of the coexistence at finite sizes.

Finally, we note that a real physical system is not expected to stay long in a state where the two phases coexist at a first-order transition. Hence, the effect of the coexistence might be less dominant for the experimental results even at finite volumes.

\subsection{Phase diagram} \label{Sec:phase_diag}

\begin{figure}
    \centering
    \includegraphics[width=0.95\linewidth]{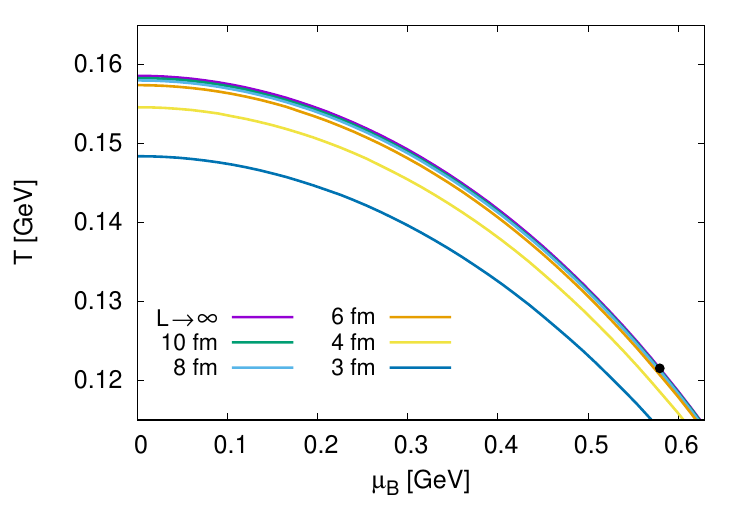}
    \caption{The chiral phase boundary obtained by the maximum of the chiral susceptibility for different system sizes using $\mathcal{U}_\eff^\infty$. The black dot marks the $L\to\infty$ CEP.}
    \label{Fig:ptByL}
\end{figure}

The size dependence of $\langle\phi\rangle$ already anticipates that the phase diagram is modified when the finite-size effects become important. At finite $L$, just as in the case of a crossover in $L\to\infty$, we determine the phase transition by the maximum of the chiral susceptibility. Using the extremum of higher-order fluctuations instead does not modify our results significantly. For simplicity and to focus on the effect of statistical weighting, first we use $\mathcal{U}_\eff^\infty$. The emerging phase boundaries---shown in Fig.~\ref{Fig:ptByL}---are surprisingly similar to the size dependence of the lattice results \cite{RubenKara:2024krv, Borsanyi:2025lim}, obtained by extrapolation from imaginary chemical potentials.
At low $\mu_{B}$, the transition temperature starts to deviate from the $L\to\infty$ value around $L=5-6$ fm with approximately $6.5~\%$ decrease at $L=3$ fm. 
At very large $\mu_{B}$, the gap between the phase boundaries at different sizes shrinks, but remains present, contrary to the lattice results.
The system size at which the phase transition deviates from the infinite-size location is governed by the weights in \Eq{Eq:Ueff_MF}.
Hence, it is sensitive to the external field $h$, which raises the $\phi<0$ and deepens the $\phi>0$ minima of the potential. 

We emphasize that including momentum-space constraints alters the finite-size effects on the phase diagram, although this effect is model-dependent. In the present approach, the discretization with PBC shifts the phase boundary to higher temperatures for decreasing $L$, reflecting the behavior of $\langle\phi\rangle$ at $\mu_B=0$. The inflection point of the potential is shifted to higher $T$ and lower $\mu_B$ since the size dependence of the fermionic vacuum term is present with full discretization. Note that the inflection point moves to lower $T$ and higher $\mu_B$ when only the zero mode of the vacuum contribution is considered, which is in line with previous results in quark-meson models \cite{Fraga:2010qef, Kovacs:2023kcn}. 
In comparison, functional methods with momentum-space constraints usually suggest the ``apparent critical point'' to be shifted to lower $T$ and higher $\mu_B$. 

Interestingly, the gradient term also moves the transition to lower temperatures (by compensating the quadratic term in the potential), while smoothening the transition by shifting the inflection point of $S_\eff^\mathrm{grad}$ to lower $T$ and higher $\mu_B$. Moreover, its effect might not be restricted to finite sizes.
When the full classical equation of motion is solved, spatially varying configurations can develop, and the interplay with the gradient term can lead to a lower total free energy compared to homogeneous configurations, as shown in~\cite{Lo:2026xuc}. This difference should be carefully assessed, as it may affect the critical endpoint even in the thermodynamic limit, and warrants further investigation.

\subsection{Thermodynamics} \label{Sec:thermodyn}

\begin{figure}
    \centering
    \includegraphics[width=0.95\linewidth]{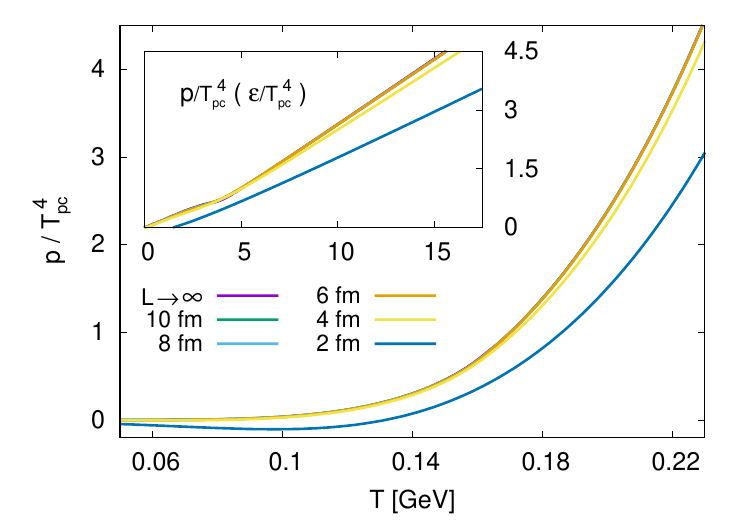}
    \caption{The pressure with the $T=0$ contribution subtracted as a function of temperature and of the energy density (inset) for different system sizes. Both $p$ and $\epsilon$ are normalized with $T_{\pc,L\to\infty}^4$ (of set C) to remove their dimensionality.} 
    \label{Fig:pressure_mu0}
\end{figure}

For completeness, we also investigate the finite-size effects on the thermodynamics at $\mu_B=0$ with $\mathcal{U}_\eff^\infty$. 
When calculating the pressure at finite $L$ and nonzero $T$, the integration over $\phi$ always has a nonzero contribution from $\phi\neq\phi_0$ with $p(\phi\neq\phi_0)<p(\phi_{0})$, and thus the pressure decreases.
We note that since at $T\to0$ the finite-size effects from the weighting are absent, a nonphysical behavior may arise around $T=0$. However, this problem is limited to very low temperatures unless the size is extremely small $L\lesssim2$ fm.
Similarly to the phase boundary, the thermodynamics does not change reasonably until $L\approx10$ fm is reached, as shown in \Fig{Fig:pressure_mu0} for $\mu_{B}=0$. 
When the finite-size effects become important, the pressure (and similarly the entropy density) decreases for the reasons mentioned above. The different contributions to the energy density $\epsilon$ compensate for each other in such a way that it remains relatively constant, with only a slight increase at higher temperatures. Consequently, the $p(\epsilon)$ curve moves downward and becomes softer with decreasing $L$, which is also shown in \Fig{Fig:pressure_mu0}. A similar behavior can be found in the vicinity of the CEP and at the first-order transition. The latter case is interesting as the pressure interpolates between the branches of the two stable mean-field solutions. However, this interpolation is always below the $L\to\infty$ pressure, just as in the $\mu_{B}=0$ case. 

The discretization with PBC in our model further decreases the pressure with the decreasing system size and makes the $p(\epsilon)$ curve softer, similar to previous results with different momentum-space constraints \cite{Kovacs:2023kbv}. On the contrary, the gradient term increases $p$, since it gives an increasing contribution to the effective action, and stiffens the equation of state.

\section{Fluctuations at finite sizes} \label{sec:Fluct}

In this section, we discuss finite-size effects on fluctuations along selected curves in the phase diagram, in particular, the phase boundary and the freeze-out line. The former provides a natural setting for theoretical analysis, while the latter is motivated by its relevance for heavy-ion phenomenology and experiments.

\subsection{Fluctuations along the phase boundary} \label{Sec:along_pb}

\begin{figure}[b]
    \centering
    \includegraphics[width=0.95\linewidth]{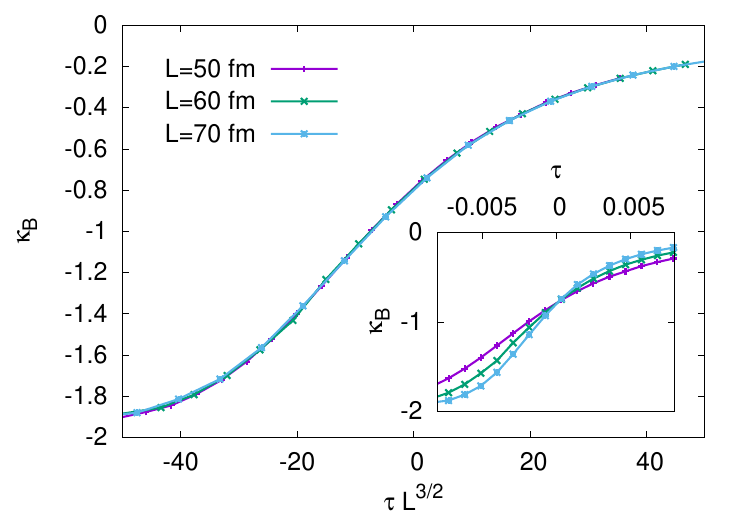}
    \caption{The Binder cumulant along the phase boundary (here determined by the minimum of $\kappa_\mathrm{B}$) as a function of $\tau L^{1/\tilde{\nu}}$ for different sizes. The inset shows the same curves but as a function of the unscaled $\tau$.}
    \label{Fig:kB}
\end{figure}

As we discussed in \Sec{Sec:chiral}, the extrema of the different order susceptibilities monotonically increase along the phase boundary, due to the scaling at the second- and first-order phase transition. Accordingly, to study the vicinity of the CEP, one needs a ratio of susceptibilities (or cumulants) that is size independent. To this end, one can define the Binder cumulant (of the net baryon number; that of the chiral susceptibility has the same qualitative behavior)
\be 
\kappa_\mathrm{B}=\frac{\kappa^B_4}{ (\kappa^B_2)^2} = \frac{\chi^B_4}{VT^3 (\chi^B_2)^2} \,,
\ee
which scales as $L^0$ both at the second-order and first-order transition \cite{Binder:1981sa}.  
To study the Binder cumulant, it is convenient to define the phase boundary by the minimum of $\kappa_\mathrm{B}(T)$. Since their effect is negligible in the discussed size range, here we omit the discretization and the gradient term and use only $\mathcal{U}_\eff^\infty$.
As shown in \Fig{Fig:kB}, 
$\kappa_\mathrm{B}$ is independent of the system size and extrapolates between the Gaussian result $\kappa_\mathrm{B}\to0$ of the crossover and $\kappa_\mathrm{B}\to-2$ obtained from a two-peaked probability approximation of the first-order case. In the close vicinity of the CEP at fixed $h$, we approximate $\tau$ by $\tau=\mathrm{sgn}(\delta T)\left(\delta T^2+\delta \mu_B^2\right)^{1/2}$ with $\delta T=T-T^\CEP$ and $\delta\mu_B=\mu_B-\mu_B^\CEP$. If $\tau$ is not rescaled with the finite size, the $\kappa_\mathrm{B}$ curves calculated at different system sizes are expected to cross each other at the critical point, which can also be used to identify the CEP when the thermodynamic limit cannot be reached \cite{Binder:1985xkp, Ashikawa:2024njc, Wada:2025ycz}. In our case, this method recovers the $L\to\infty$ CEP as expected, shown in the inset of \Fig{Fig:kB}.

From the above discussion, it is clear that the ratio $\chi_4^B/\chi_2^B$---used also in Refs.~\cite{Kovacs:2023kbv} and \cite{Bernhardt:2023hpr}---is appropriate outside of the scaling region, but should be volume dependent in the vicinity of the CEP. This is in line with the finite-size scaling that predicts $\chi^B_4/\chi^B_2\propto L^{2 y_h}=L^{9/2}$ near the critical point, in contrast to $\chi^B_4/\chi^B_2\propto L^0$ far from it, for instance, at the freeze-out line.

\subsection{Fluctuations along the freeze-out line} \label{Sec:along_fo}

\begin{figure}
    \centering
    \includegraphics[width=0.95\linewidth]{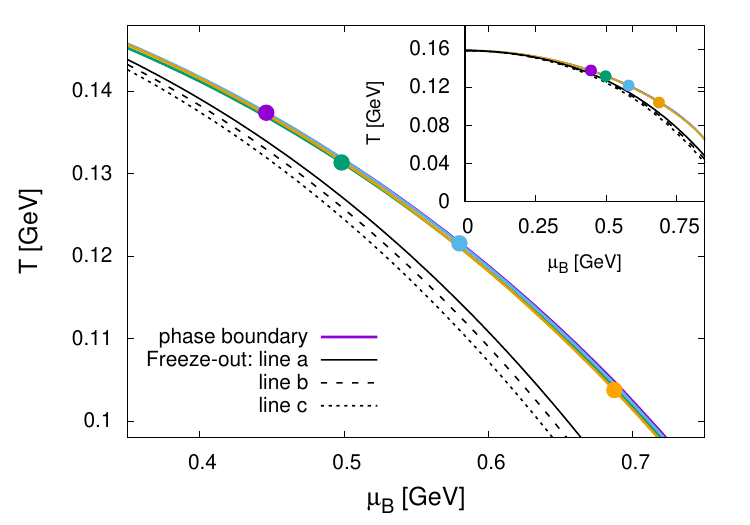}
    \caption{The phase boundaries with the $L\to\infty$ CEP for the parameterizations A-D in Tab.~\ref{tab:param_sets} and the three freeze-out line parametrization we use.}
    \label{Fig:freezeouts}
\end{figure}

\begin{figure}
    \centering
    \includegraphics[width=0.95\linewidth]{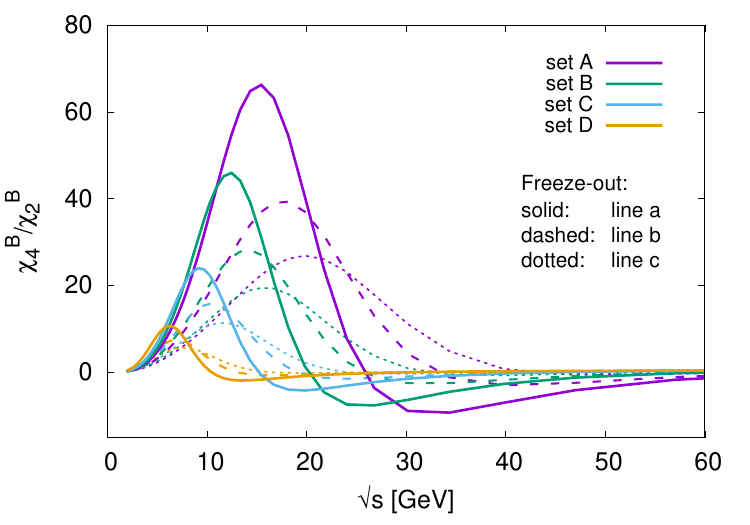}
    \caption{The $R_{42}$ baryon number cumulant ratio along the freeze-out lines shown on \Fig{Fig:freezeouts} calculated with the different parameterizations at infinite size using $\mathcal{U}_\eff^\infty$.}
    \label{Fig:R42_Vinf}
\end{figure}

\begin{figure}
    \centering
    \includegraphics[width=0.95\linewidth]{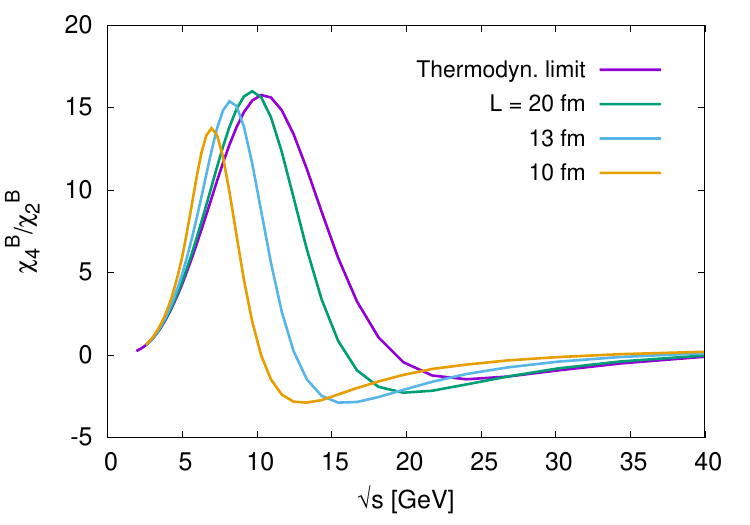}
    \caption{The $\chi_4^B/\chi_2^B$ susceptibility ratio at different sizes along the freeze-out line b using $\mathcal{U}_\eff^\infty$ and the parameter set C.}
    \label{Fig:R42_freezeout}
\end{figure}

For the experimental relevance of the finite-size effects, we investigate the fluctuations also along the freeze-out line, where non-monotonicity of the cumulant ratios is widely discussed in the context of the critical endpoint \cite{Bzdak:2019pkr, PhysRevLett.126.092301, STAR:2022etb, HADES:2020wpc}. Since this line is outside of the scaling region, one expects $\kappa_k\propto L^d$ for the $k^\mathrm{th}$ cumulant and hence $\chi_k\propto L^0$ for the corresponding susceptibility. Therefore, one may use the cumulant or susceptibility ratios $\kappa_k/\kappa_l\sim\chi_k/\chi_l$ with $l,k\neq l\in\mathbb{Z}^+$, which behave as $\propto L^0$ when the genuine finite-size effects are not yet present. 
Moreover, instead of using the temperature or baryon chemical potential, these parameters can be related to the $\sqrt s$ center of mass energy according to the parameterization in Ref.~\cite{Andronic:2017pug}. The freeze-out line itself can be directly obtained from $T(\sqrt{s})$ and $\mu_{B}(\sqrt{s})$ as
\be \label{Eq:Fo_Andronic}
T_\fo (\mu_{B}) = T_\fo^0/\left(1+\exp{a - b \ln{(\mu_{B}^0/\mu_{B} - 1)}}\right) \,,
\ee 
or we can use a simple quartic form
\be \label{Eq:Fo_Clemens}
T_\fo (\mu_{B}) = T_\fo^0 - \alpha \mu_{B}^2 - \beta \mu_{B}^4 \,,
\ee 
with $\alpha,\beta>0$ as in Ref.~\cite{Cleymans:2006qe}.
Since the definition in Eq.~\eqref{Eq:Fo_Andronic}, being parametrized to obtain the $\sqrt{s} - \mu_B, T$ relation, is not completely compatible with the phase boundary in our model, we will employ \Eq{Eq:Fo_Clemens} with setting $T_\fo^0=T_\pc|_{\mu_{B}=0}$.
Our description is only of a qualitative nature, therefore we aim to find a setup which is qualitatively sensible. Given uncertainties in both the exact location of the freeze-out line and the phase boundary at large chemical potential, we try multiple parameterizations of Eq.~\eqref{Eq:Fo_Clemens}. Fixing $\beta=0.06$, we use $\alpha=0.110$, $0.115$, and $0.120$ labeled as lines $a$, $b$, and $c$, respectively. To test the effect of CEP at different locations, we also use multiple parameter sets (see \Tab{tab:param_sets}). The obtained phase transition lines are (almost) on top of each other, while the CEP is shifted as shown in \Fig{Fig:freezeouts}, where the freeze-out lines are also depicted. 
Here we use $\mathcal{U}_\eff^\infty$ and briefly discuss the effect of momentum space discretization and the gradient term subsequently.
We show the "kurtosis" of the net baryon number $R_{42}=\chi_4^B/\chi_2^B$ calculated along the different freeze-out lines with the different parameterizations at $L\to\infty$ in \Fig{Fig:R42_Vinf}.
As the relative separation between the freeze-out curve and the phase boundary becomes wider, the amplitude of the signal decreases significantly. This demonstrates the importance of the distance of the phase boundary and the CEP from the freeze-out. The absolute $\sqrt{s}$ dependence and the location of the nonmonotonicity are model dependent and rely on the nonuniversal mapping to Ising variables. Therefore, direct quantitative comparison of $\sqrt{s}$ with values in the experimentally relevant region $\sqrt{s}\lesssim 10$ GeV is not intended.
To discuss the finite-size dependence, we consider line $b$ and parameter set C. While the specific choice for the freeze-out line affects the magnitude of the fluctuations, it does not alter our findings qualitatively. 
The obtained behavior is shown in \Fig{Fig:R42_freezeout}.
The infinite volume dip-peak structure persists with moderate suppression even for $L\approx 10~\text{fm}$, but is shifted to lower $\sqrt{s}$ for sufficiently small $L$. While the phase transition exhibits only weak size-dependence above $ L\approx10$ fm, the fluctuation on the freeze-out remains sensitive up to $L\approx40$ fm, since even a small shift in the relative position of the phase boundary and the freeze-out line can significantly impact observables on the latter. For linear sizes $L\gtrsim 40~\text{fm}$ there is no finite-size effect, and practically, the $L\to\infty$ behavior is seen.
However, the system sizes typical for experiments, especially at lower energies, are clearly within the range where the finite-volume effects can even significantly alter the observables.

\begin{figure}
    \centering
    \includegraphics[width=0.95\linewidth]{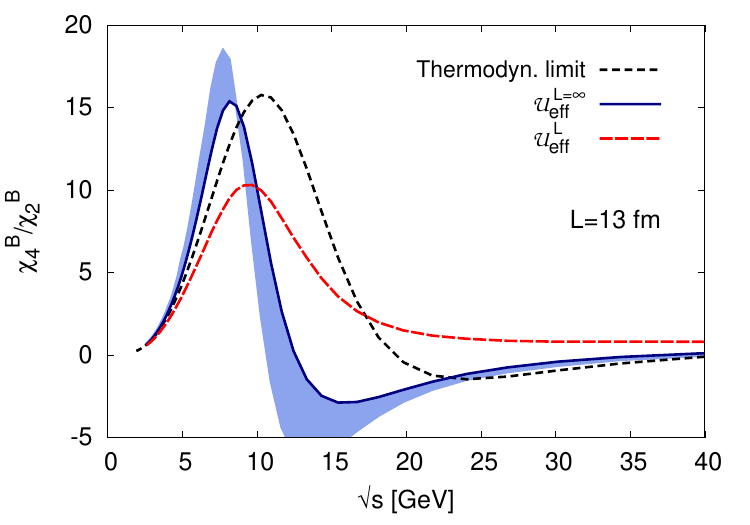}
    \caption{The effect of the modeled gradient term for $c\leq6$ (represented by the shaded area) and the discretization of fermion fluctuations on the $\chi_4^B/\chi_2^B$ susceptibility ratio at $L=13$ fm. The result in the thermodynamic limit is shown for comparison.}
    \label{Fig:R42_freezeout_models}
\end{figure}

Including momentum-space discretization modifies the effective potential and, consequently, the resulting phase structure and fluctuations. Its impact is primarily determined by the direction in which it shifts the phase boundary and the inflection point of the potential, hence raising significant model dependence. In our framework, employing $\mathcal{U}_\eff^{L}$ with PBC increases the transition temperature while pushing the inflection point to lower chemical potentials. This compensates for the shift but leads to a stronger damping of the dip-peak structure as shown in Fig.~\ref{Fig:R42_freezeout_models}. The modeled gradient term has an opposite effect on the phase boundary, therefore it enhances the finite-size shift of the nonmonotonicity when taken into account. This trend is demonstrated by the shaded region in Fig.~\ref{Fig:R42_freezeout_models}, which is obtained by varying the weight of the gradient term in the ansatz in Eq.~\eqref{Eq:Skin} over a prescribed range.
In conclusion, the momentum-space constraints and the gradient term alter but do not eliminate the finite-size effects.

We emphasize that the nonmonotonic structure in \Fig{Fig:R42_freezeout} is not an unambiguous signal of a nearby critical point, especially when the coexistence gives such a dominant contribution as in our approach. 
Recall that at finite system sizes, the extremum values of the susceptibilities and hence of $R_{42}$ show a monotonic increase/decrease with increasing $\mu_B$. In such a finite system, there is no unique signature of the $L\to\infty$ critical endpoint in $R_{42}$, or in $\chi_4$ and $\chi_2$ individually, if only a single volume is considered. 
The dip-peak structure along the freeze-out line arises simply from the separation of the phase boundary and the freeze-out. Therefore, it can not identify the vicinity of an $L\to\infty$ critical point at finite sizes unless further information is known, e.g., the quantitative value of the fluctuations with also finite-size effects included.

Finally, we note the existence of an intermediate region between the phase boundary discussed in \Sec{Sec:along_pb} and the freeze-out line discussed here. Since in a given $L$ range, the extent of the finite-size scaling region for $\chi_4$ turns out to be larger than that for $\chi_2$, there is a part of the phase diagram, where the former still shows critical $L$-scaling, while the latter is already $\propto L^0$. This gives rise to a peculiar scaling behavior that cannot be described by a constant $\chi_4/\chi_2$ or $\kappa_\mathrm{B}$. However, not only the size and location, but even the existence of this interesting region is clearly model-dependent. 

\section{Conclusion} \label{sec:conclusion}

We studied the finite-size effects in a schematic quark-meson type model based on the size dependence of the partition function and the free energy, defined via a nonperturbative integration over the constant field mode. We mainly used an approximation based on a volume-independent effective potential. In this approach, we found that the finite-size effects decrease the (pseudo)critical temperature and soften the chiral transition.
We were able to recover the correct finite-size scaling behavior at the critical point, while seeing an even stronger volume dependence of the fluctuations at the first-order transition caused by the phase coexistence. From the scaling of the different susceptibilities, $\chi_k$, $\chi_k^B$, $\chi_k^T$, it follows that each $h$, $\mu_{B}$, and $T$ has an overlap with the Ising external field-like scaling direction. This can also be seen within more complex models (e.g., advanced versions of the quark-meson model \cite{Kovacs:2016juc}) by studying the critical scaling at $L\to\infty$.

We discussed that momentum-space constraints can alter the finite-size dependence, leading to an increasing pseudocritical temperature. However, their effect can vary greatly depending on the boundary condition and the treatment of the vacuum term \cite{Kovacs:2023kbv}. 

We also incorporated the effect of the gradient term, which is rarely addressed in mean-field studies, and find that it provides a substantial contribution below system sizes $L \approx 10-20$ fm, leading to a significant reduction of the pseudo-critical temperature and a softening of the transition. 
This indicates that gradient effects are not merely subleading in this regime, and that a quantitatively controlled description will require a self-consistent treatment of spatially varying field configurations, obtained by solving the corresponding equation of motion and determining the resulting profile, along the lines of \cite{Lo:2026xuc}; this is left for future work.

Since showing a monotonic behavior along the phase boundary due to the coexistence, no apparent critical endpoint can be identified by the extremum of the susceptibilities. This behavior is expected for the properly defined susceptibility \cite{Binder:1992pz}, but contrasts models incorporating finite-size effects only via momentum-space constraints in functional approaches \cite{Almasi:2016zqf, Tripolt:2013zfa, Bernhardt:2021iql}. The $L\to\infty$ critical point can be located using the Binder cumulant, which we discussed along the phase boundary.

Along the freeze-out line, which lies completely outside the scaling region in our scenarios, we studied the kurtosis ratio $R_{42}=\chi_4^B/\chi_2^B$. In a finite system, the dip-peak structure of $R_{42}$ along the freeze-out line can be associated with merely the deviation of the freeze-out and the phase transition. Therefore, the non-monotonic behavior does not necessarily indicate the proximity of a critical endpoint.
\red{$R_{42}$} is more sensitive to the finite-size effects, therefore the emerging dip-peak structure is suppressed and shifted to higher chemical potentials, hence lower $\sqrt{s}$, already at $L\lesssim 40$ fm. Momentum discretization and the modeled gradient term alter but do not eliminate the finite-volume modifications. The results demonstrate that the finite-size effects can be relevant in the size regime typical for experiments, especially at lower energies. This highlights the need for a comprehensive theoretical description, including finite-size effects, to attempt locating the critical endpoint through the baryon fluctuations by comparison to experimental results.

Although our schematic framework is capable only of qualitative analyses, it provides the possibility of discussing other physical quantities at finite volume. For instance, it can be used to study not only a Lee-Yang edge singularity, but also Lee-Yang zeros in the vicinity of the critical point \cite{Wada:2024qsk, Adam:2025phc}, and the interplay of different special points in the space of complex parameters. 
It would also be interesting to directly compare the finite-size effects near a critical point with lattice calculations, which can be performed in heavy-quark QCD \cite{Kiyohara:2021smr, Ashikawa:2024njc, Wada:2024qsk}.
Finally, while our approach is built on the grand canonical picture, a canonical ensemble is expected to be more realistic for very small systems \cite{Hagedorn:1984uy, Cleymans:1998yb, Braun-Munzinger:2020jbk, Friman:2025swg}. The canonical effects and their relation to our result might also be addressed in future work.

\acknowledgements

We thank Zsolt Szép, Attila Pásztor, Masakiyo Kitazawa, Tatsuya Wada, and Takahiro Doi for insightful discussions.
We are grateful for the comments and discussion during the INT workshop INT-25-3a. 
This work is partially supported by the Polish National Science Centre (NCN) under OPUS Grant No. 2022/45/B/ST2/01527. 
G.~K. acknowledges support from the Polish National Science Centre, Poland (NCN), MINIATURA 9, grant No. 2025/09/X/ST2/00748, and from the Hungarian National Research, Development and Innovation Fund under Project number K 138277.
K.R. acknowledges the support of the Polish Ministry of Science and Higher Education. 
C.S. acknowledges the support of the World Premier International Research Center Initiative (WPI) under MEXT, Japan.

\appendix

\section{Thermodynamic quantities at finite size} \label{sec:defs}

For completeness, here we define the physical quantities in the presented framework derived from the free energy in \Eq{Eq:freeenergy_def}, which also serves as a cumulant generating function. We focus on the case of a constant background field and size-independent effective potential; extension to the size-dependent case is straightforward.
We start the definitions with the chiral quantities that can be obtained by derivatives with respect to $h$ at fixed $T$, $\mu$, and $V$. The chiral condensate reads (we keep the $S=\beta V\mathcal{U}_\eff$ in the exponents to ease the notation)
\be \label{Eq:phi_avg_formal} \begin{split}
\langle \phi \rangle =& -\frac{1}{V}\frac{\delta \Phi}{\delta h} =
\frac{1}{\mathcal{Z}} \int d \phi~\phi~ e^{-S} 
\,,
\end{split}
\ee 
while the chiral susceptibilities (we use $\chi_1\equiv\langle\phi\rangle$, $\chi_2\equiv\chi$ in the main text) are
\be \label{Eq:chi_formal} \begin{split}
\chi_2 =& \frac{\delta \langle \phi \rangle}{\delta h} = \frac{1}{\beta V} \left( \frac{1}{\mathcal{Z}} \frac{\delta^2 \mathcal{Z}}{\delta h^2} - \left(\frac{1}{\mathcal{Z}} \frac{\delta \mathcal{Z}}{\delta h}\right)^2 \right) \\
=& \beta V \left( \langle \phi^2 \rangle - \langle \phi \rangle^2 \right) \,,\\
\chi_3 =& (\beta V)^2 \left( \langle \phi^3 \rangle - 3  \langle \phi^2 \rangle \langle \phi\rangle +2\langle\phi\rangle^3\right)\\
\chi_4 =& (\beta V)^3 \big(\langle\phi^4\rangle - 4 \langle\phi^3\rangle\langle\phi\rangle \\
& \qquad + 12 \langle\phi^2\rangle\langle\phi\rangle^2-3\langle\phi^2\rangle^2-6\langle\phi\rangle^4\big) \,.
\end{split} \ee 
where $\langle\phi^k\rangle$ is the $k^\mathrm{th}$ raw moment. These susceptibilities are directly related to the cumulants, while their relation to the central moment is as usual. 

The thermodynamic quantities can be defined analogously as in the grand canonical ensemble with 
\be 
\Phi = \langle E \rangle - T \mathrm{S} -\mu_B \langle N\rangle
\ee 
and
\be 
d\Phi = - \mathrm{S} dT - P dV - \langle N\rangle d\mu_{B}
\ee 
Accordingly, the expectation value of the baryon number (particle number, in general) is obtained as the derivative with respect to $\mu_{B}$ with all other parameters ($V$, $\beta$, $h$) fixed
\be \label{Eq:N_avg_def} \begin{split} 
\langle N \rangle =& -\frac{\partial\Phi}{\partial\mu} =  -\frac{V}{\mathcal{Z}} \int d \phi ~\dot{\mathcal{U}}_\eff \, e^{-S}\,,
\end{split} \ee
where $\dot{f}\equiv\partial f/\partial\mu_{B}$. 
The cumulants of the baryon number can be obtained by
\be \begin{split}
    \kappa_k = \frac{\partial^{k} \log \mathcal{Z}}{\partial (\beta\mu_B)^{k}}= \frac{\partial^{k-1} \langle N\rangle}{\partial (\beta\mu_{B})^{k-1}} 
\end{split} \ee
for $k\geq1$ (for $k=1$, $\kappa_1=\langle N\rangle$). 
To remove the extensive volume dependence and allow direct comparison to infinite volume results, in the calculations we use the global baryon number density $n=\langle N\rangle /V$ and the generalized susceptibilities $\chi^B_k=\kappa_k/(VT^3)$. For the fluctuations, one can also use the cumulant densities $\kappa_k/V$, however, $\chi^B_k$ is more widely used in the literature.

In \Eq{Eq:N_avg_def}, one might identify the baryon number for a given field configuration to be $\mathbf{N}(\phi)=-V\,\dot{\mathcal{U}}_\eff$. 
Interestingly, the second raw moment of the baryon number (using $\mathcal{Z}$ to define the moment generating function as $M(\lambda)=\mathcal{Z}(\mu_{B}+\lambda/\beta)/\mathcal{Z}(\mu_{B})$ and $\langle N^k \rangle_{\mathrm{raw}}=\partial ^{k}M(\lambda)/\partial\lambda^{k}|_{\lambda=0}$, where $\langle\cdot\rangle_{\mathrm{raw}}$ stands for the raw moment and not for the averaging) reads
\be \begin{split}
 \langle N^2 \rangle_\mathrm{raw} =& \frac{1}{\mathcal{Z}} \int d \phi (V^2\dot{\mathcal{U}}_\eff^2 - TV\,\ddot{\mathcal{U}}_\eff) e^{-S} \\
 =& \langle \mathbf{ N}^2 \rangle + T\langle \dot{\mathbf{ N}} \rangle \,. 
\end{split}
\ee 
Here, the second term corresponds to the chemical potential dependence of the particle number corresponding to a given field configuration. 
This emerges since we calculate expectation values as an average over the possible field configurations (now trivially, the value of $\phi$) instead of the microstates with given $N$. Calculating the second raw moment density $\langle N^2\rangle_\mathrm{raw}/V^2$, one can see that 
\be
\langle N^2\rangle_\mathrm{raw}/V^2 = \langle \dot{\mathcal{U}}_\eff^2 \rangle + T/V \, \langle - \ddot{\mathcal{U}}_\eff \rangle \,,
\ee 
i.e., the extra term is suppressed for large volumes and small temperatures, as expected.

We emphasize that the baryon number fluctuations come as the derivative of $\Phi$, and not as a derivative of the pressure, which is defined as the derivative of the free energy with respect to volume
\be \label{Eq:press_def} \begin{split}
P\equiv-\frac{\partial \Phi}{\partial V} =  -\frac{1}{\mathcal{Z}} \int d\phi ~\mathcal{U}_\eff\, e^{-S }\,
\end{split} \ee 
and hence, $P\neq-\Phi/V$. Generally, division by extensive variables such as the volume in the grand canonical ensemble and also the particle number in the canonical ensemble is often used to remove the extensivity of other quantities. However, this is not equivalent to calculating the corresponding derivatives. Moreover, we note that the derivations with respect to different parameters, $V$, $\mu_{B}$, $\beta$, and $h$ (with all other parameters kept fixed) are not interchangeable with each other and with the averaging~$\langle\,\cdot\,\rangle$. (It is also clear that the quantities derived from the grand free energy depend on all those parameters, unless a coincidental cancellation happens.) We note that for convenience, and to distinguish from the $P(\phi)$ probability distribution \Eq{Eq:phi_def}, we denote the pressure with lowercase $p$ in the main text.

For completeness, we also define the entropy 
\be \label{Eq:Entropy_def}
\mathrm{S}= -\frac{\partial \Phi}{\partial T} =- \frac{ \Phi}{T} + \frac{1}{T} \frac{\partial (-\ln \mathcal{Z})}{\partial \beta}
\ee
and the energy 
\be \begin{split}
\langle E\rangle= \frac{\partial(-\ln \mathcal{Z})}{\partial\beta} - \frac{\mu_{B}}{\beta} \frac{\partial(-\ln \mathcal{Z})}{\partial\mu_{B}}=\Phi + T\mathrm{S}+\mu_{B} \langle N\rangle 
\end{split} \ee 
as well. To remove the extensivity, for simplicity, we define the corresponding global entropy and energy densities (energy and entropy per unit volume) as $s=\mathrm{S}/V$ and $\epsilon=\langle E\rangle/V$, respectively.

\bibliography{biblio}

\end{document}